\documentclass[12pt]{spieman}  
\usepackage{amsmath,amsfonts,amssymb}
\usepackage{graphicx}
\usepackage{setspace}
\usepackage{tocloft}
\usepackage[dvipsnames]{xcolor}
\usepackage{xcolor}

\title{Demonstration of Roman Coronagraph Instrument Star Acquisition in Thermal Vacuum.\footnote{This paper has been published in the Journal of Astronomical Telescopes, Instruments, and Systems (JATIS), VOL. 11 · NO. 2 | April 2025, SPIE Digital Library.}}
\author[a]{Nanaz Fathpour}
\author[a]{Milan Mandić}
\author[a]{Joel Shields}
\author[a]{Zahidul Rahman}
\author[a]{Alfredo Valverde}

\affil[a]{Jet Propulsion Laboratory, California Institute of Technology, 4800 Oak Grove Drive, Pasadena, CA, USA 91109}

\cftpagenumbersoff{figure}
\cftpagenumbersoff{table} 

\usepackage{lipsum}

\newcommand\blfootnote[1]{%
  \begingroup
  \renewcommand\thefootnote{}\footnote{#1}%
  \addtocounter{footnote}{-1}%
  \endgroup
}

\begin{document} 
\maketitle

\begin{abstract}
NASA’s Nancy Grace Roman Space Telescope is a flagship astrophysics mission planned for launch  no later than May 2027. The Coronagraph Instrument (CGI) aboard Roman will demonstrate the technology for direct imaging and spectroscopy of exoplanets around nearby stars and starlight suppression that surpass previous space-based and ground-based coronagraphs.  This is accomplished using active wavefront control in space with deformable mirrors.  CGI requires a star acquisition system capable of acquiring reference and target stars to achieve this goal.  This paper describes two CGI star acquisition system methods: EXCAM and Raster Scan.  Furthermore, it will detail the results of the CGI thermal vacuum (TVAC) tests conducted to evaluate system-level star acquisition and verify its performance requirements.  
\end{abstract}


{\noindent \footnotesize\textbf Email:  \linkable{nanazf@gmail.com}, \linkable{milan.mandic@jpl.nasa.gov} }

\blfootnote{\copyright{} 2024 California Institute of Technology. Government sponsorship acknowledged.}
\begin{spacing}{1}   

\section{Introduction}
\label{sect:intro}  
The Nancy Grace Roman Space Telescope (RST)  is a NASA observatory designed to settle essential questions in dark energy, exoplanets, and infrared astrophysics. The Roman Space Telescope, managed by NASA's Goddard Space Flight Center (GSFC), is designed for a five-year mission and will launch from Cape Canaveral to L2 no later than May 2027. The telescope will feature two instruments: the Wide Field Instrument (WFI) and the Coronagraph Instrument (CGI).

The Coronagraph Instrument on the Roman Space Telescope is an advanced technology demonstration, paving the way for future missions.  It is designed to conduct high-contrast imaging and spectroscopy of dozens of individual nearby exoplanets.  Developed by NASA’s Jet Propulsion Laboratory (JPL), the Roman Coronagraph Instrument will be instrumental in showcasing the necessary technologies for future missions aimed at imaging and characterizing rocky planets within the habitable zones of nearby stars. By testing these advanced tools through an integrated end-to-end system and conducting scientific observation operations, NASA will validate performance models and lay the groundwork for potential flagship missions, such as the Habitable World Observatory (HWO).

CGI is a technology demonstration (tech demo) showcasing key hardware elements and algorithms for star acquisition, active wavefront control, sensing, pointing control and analysis, and data post-processing. For more information on the CGI, refer to these papers\cite{IP2020, IP2024,JK2023} in the reference.
A vital aspect of flying CGI will be understanding how the coronagraph interacts with the entire observatory and its spacecraft control systems, thereby significantly reducing the risks for future missions.  

Successful star acquisition is among the initial and critical tasks CGI must undertake during its operational phase. Depending on mission phases, CGI performs star acquisition using two different methods.  Both methods require the RST Attitude Control System (ACS) to provide supplementary pointing capabilities.

The first method is the EXCAM star acquisition which happens at the beginning of an observation and uses Exoplanetary System Camera (EXCAM).  This is during initial acquisition when the detector's Field of View (FoV) is not obscured since the Focal Plane Mask (FPM) has not yet been inserted into the optical path.

The second method is Raster Scan Star Acquisition, which employs a Fast Steering Mirror (FSM) in conjunction with a Low Order Wavefront Sensing (LOWFS) Camera, LOCAM.   At this stage, the FPM has been inserted, and dark hole (DH) creation or starlight suppression has been successfully achieved.  Therefore, we can't use the EXCAM star acquisition method since EXCAM FoV is now obscured, and removing masks can degrade or destroy the high-contrast state and affect the quality of the resulting science data. 

This paper provides a brief overview of the star acquisition system integrated within the CGI, detailing its architectural design and presenting results from TVAC testing conducted at JPL.

\section{CGI Star Acquisition Overview}
\label{sec:overview}

In this section, we will give an overview of the CGI star acquisition system and two different methods of star acquisition depending on mission phases.  

The first method is EXCAM star acquisition, which uses EXCAM, CGI's science camera, to acquire a star. This requires the spacecraft (S/C) ACS to bring the star of interest to the EXCAM FoV.  CGI will perform EXCAM star acquisition at the beginning of the CGI observation. 
 
 After operations with the WFI, RST’s primary instrument, a Survey Observations Sequence (SOS) package, will be transmitted from the ground, outlining CGI’s next bright reference and target stars.  A reference star serves the purpose of a bright star and, in some instances, will be the 
 same, but its main objective is to provide an update to the dark hole state and a reference image
 for data post-processing. This SOS will designate desired pixel locations for each WFI science 
 camera array (SCA) corresponding to the selected bright reference and target stars. The pixel
  locations will be carefully chosen so that as the WFI, functioning as a fine guidance sensor (FGS),
 stabilizes on these designated stars, CGI’s star of interest will fall within the detector’s field of view.

 The EXCAM field of view (FoV) is $19.5"\times19.5"$. During the initial CGI star acquisition, the ACS system places the desired CGI star within a 4" radius of the EXCAM's center pixel per interface requirements between the CGI and ACS systems. Now that the star is within the EXCAM FoV, the CGI can directly acquire it using its algorithms.
  
 After successful EXCAM star acquisition process and achieving the required contrast, the FPM is inserted and aligned into the optical path.  The optical path must remain undisturbed
 for the rest of the observation. Since the FPM is in the optical path, it blocks most of the 
 EXCAM FoV.  Therefore, EXCAM star acquisition is no longer feasible.

In this phase, the CGI science or tech demo observations start, and the raster scan star
 acquisition method will be used to acquire stars.  During the raster scan star acquisition, the ACS
system places the desired star within a 0.3" radius of the FPM's center per another interface
requirement between the CGI and the ACS systems.  The observatory maneuvers from bright
reference stars to dim target stars with slews and rolls.  During this phase, the FSM rasters the
sky while LOCAM records intensity counts as a function of FSM location.  The scanning motion
of the FSM and the intensity data collected by the LOCAM are used to find the star in the
LOCAM image.

\begin{figure}
    \centering
    \includegraphics[width=0.85\linewidth]{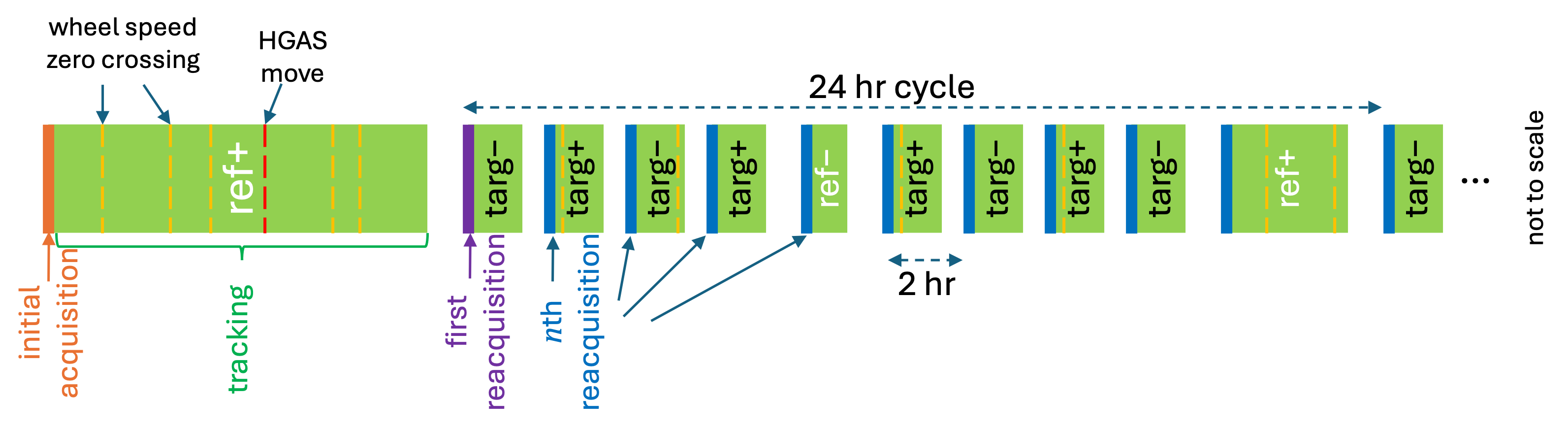}
    \caption{An example of the CGI observing scenario and star acquisition during initial acquisition
    of a reference star and then acquisition and reacquisition of target and reference stars after
     every slew/roll.}
    \label{fig:os11}
\end{figure}

An observing scenario for the CGI is given in Fig.~\ref{fig:os11},  where initial star acquisition happens at the beginning of the observation using the EXCAM star acquisition method.  After some calibration events, the dark hole is dug on the reference star, meaning the required contrast has been achieved for that reference star.  At this point, the CGI tech demo starts when the acquisition and reacquisition of the target star happen after each slew or roll using the raster scan star acquisition method.  
The CGI uses the ACS pointing capability through offloads, as explained in the following sections. 

CGI uses Reference Differential Imaging (RDI) and Angular Differential Imaging (ADI) techniques in data post-processing to produce better imaging contrast. 
The slews occur during the transition from reference stars to target stars, and vice versa, as part of the  RDI technique.
The rolls occur at the target star, between two roll attitudes, as part of the ADI technique\cite{JK2023}. Therefore, the target star would be acquired and re-acquired during rolls as shown in Fig.~\ref{fig:os11}.  It is expected that planets will orbit around target stars.

\begin{figure}
    \centering
    \includegraphics[width=0.65\linewidth]{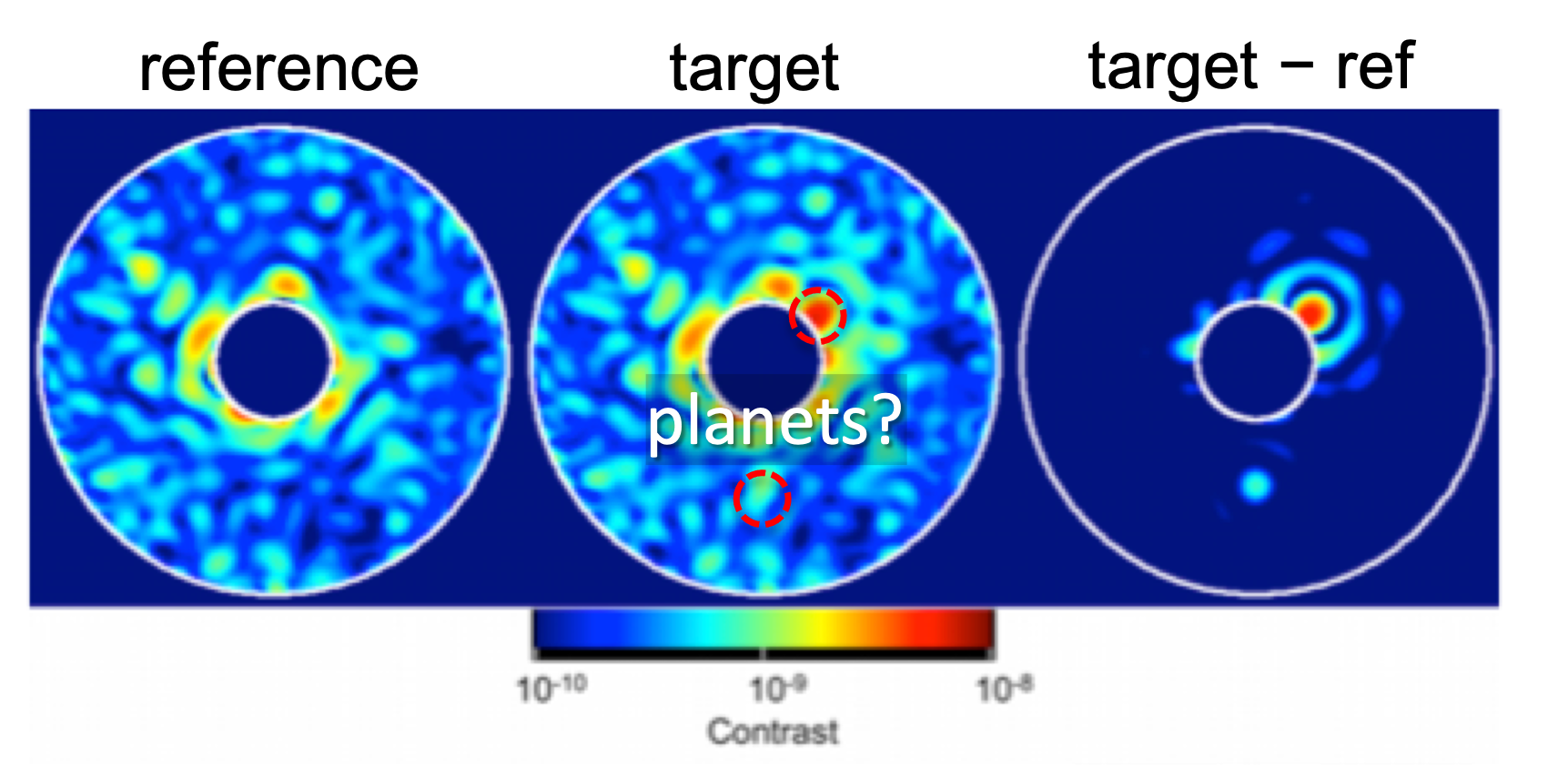}
    \caption{Reference Differential Imaging (RDI) is a post-processing technique used to remove speckles 
    from the target star by subtracting the reference star, with similar speckle patterns, to reveal 
    any planets.}
        \label{fig:rdi}
\end{figure}

 Residual starlight in the dark hole, also known as speckles, has a similar morphology to a
planet.  RDI, as shown in Fig.~\ref{fig:rdi}, subtracts these speckles to reveal planets around the target star
of interest.  First, a reference speckle pattern from another star is obtained.  The reference star
should have a similar angular size as the target star of interest.  Then speckle pattern of this
reference star would be subtracted from the target star's pattern in the post-processing to reveal
planets around these target stars.\cite{JK2023}

 The Pointing, Acquisition, and Control Element (PACE) team was responsible for the star acquisition system, and pointing, Focus, and Zernike control systems.  The team was responsible for architecting, designing, analyzing, and delivering the code for the CGI star acquisition, pointing Line of Sight (LoS) control system as well as Focus and Zernike controls\cite{NF2019,NF2022}. The team was also responsible for star acquisition and control algorithms during various JPL testing campaigns, including CGI TVAC testing.  For more information on test results for CGI control systems, see these papers\cite{MM2024,JBS2024} in reference.

The following sections will explore the two methods of CGI star acquisition, followed by a discussion on the information exchange between the CGI and Roman ACS through the offloads that CGI sends to the Roman ACS for additional pointing capability throughout the CGI star acquisition and observation phases. 
    Finally, we will report on the results of the TVAC tests, where we successfully validated the star acquisition algorithms within the CGI TVAC chamber for both star acquisition methods.
  
\subsection{EXCAM Star Acquisition Mode}
\label{sec:excam}
It is important to note that, while this paper presents star acquisition for coronagraphic observations, these algorithms will also be used for calibrations and non-coronagraphic observations during CGI operations.

\begin{figure}
    \centering
    \includegraphics[width=1.0\linewidth]{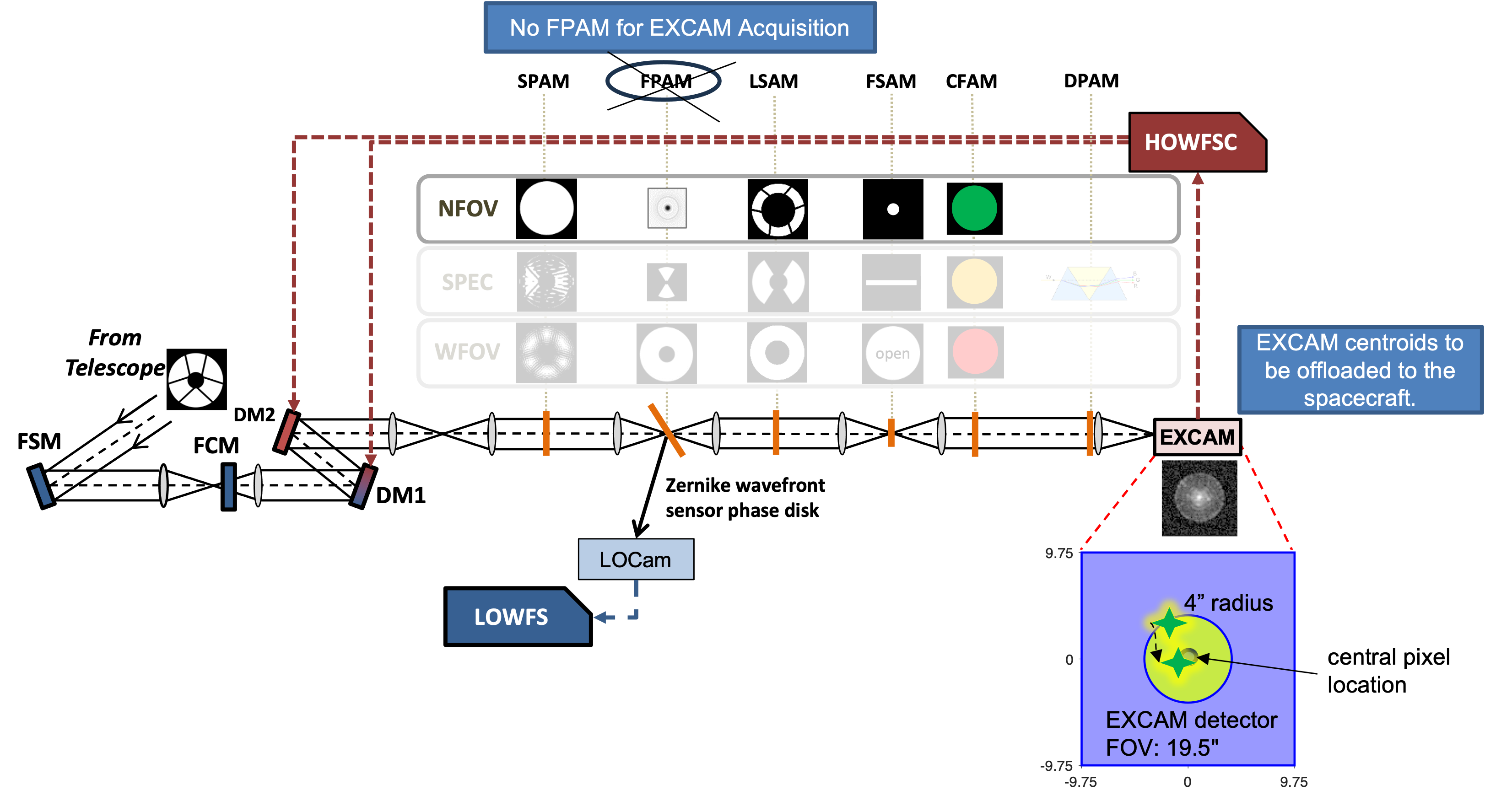}
    \caption{The CGI optical configuration during EXCAM star acquisition when the EXCAM is unobscured
     since the FPM has not been inserted yet.   The EXCAM FoV is $19.5"\times19.5" $ and the ACS would
     place the desired star within a 4" radius of the EXCAM center pixel. Since the star is now on the
     EXCAM, the CGI then computes the centroid directly and sends the star's location relative to the 
      central pixel to the ACS. ACS then adjusts pointing and places the star within the desired pixel
      location. (FoV is not to scale)}
    \label{fig:lowfs_excam}
\end{figure}

EXCAM star acquisition happens once per observation when the optical path is clear,  and the science camera FoV is unobscured. Fig.~\ref{fig:lowfs_excam} shows the optical configuration during the EXCAM star acquisition before the dark hole and when the FPM has not been inserted yet; therefore, the optical path is unobscured.  
EXCAM detector has a FoV of 19.5 $\times$ 19.5 arc-seconds.  However, the Roman
ACS can place the CGI's star of interest at a 4 arc-second radius from the EXCAM center, as shown
in Fig.~\ref{fig:lowfs_excam}.
This is an interface requirement between the CGI and the ACS systems to guarantee a successful star acquisition for the CGI.

The ability of the spacecraft ACS to bring the desired star into CGI’s FoV is limited by various errors, including a rigid body rotation calibration, star catalog errors, thermal deformations that affect the relative boresight alignment between WFI and CGI,  ACS errors, and observatory high-frequency jitter, among others.  This pointing error has been determined to be less than 4 arc-second radius, $3\sigma$.\footnote{Default acquisition  algorithm searches for a radius of over 5 arc-seconds, in case this search area needs to be expanded, there is a parameter(ACQ EX Border Margin) in the CGI flight software that can be updated.}

Before initial acquisition and taking frames, the EXCAM needs to be calibrated and configured for initial EXCAM Star Acquisition.  The EXCAM gain and exposure time must be set, and the number of frames must be selected.  Before initiating the star acquisition algorithm, the master dark must be updated to match the commanded gain and exposure time.  Then, the raw frames will be cleaned from cosmic rays and detector noise. During initial acquisition, these cleaned EXCAM frames will be used to locate the star within the frame.

An algorithm, generically called Star ID, will determine whether the star of interest is in the image using a photometric threshold – meaning that for identification of the star, the peak amplitude must be higher than a predetermined value.  This threshold is a parameter in the flight software and can be updated if necessary.  
Since we are ensured that any star within 12 arc-seconds (on-sky) of our star of interest is at least three visual magnitudes dimmer, selecting this threshold is not considered a challenge. 

For the CGI tech demo star acquisition scenario, It is assumed that there cannot be stars with
 similar brightness in the detector's FoV.  All stars of interest would have a visual magnitude of
less or equal to 5 ($V_{mag} \leq 5$), and there is a low probability that there would be a star brighter than
 $V_{mag} = 8$ in the EXCAM FoV.  It is important to note that to guarantee a successful EXCAM star 
acquisition, it is expected that when targeting a star with visual magnitude  $V_{mag}$\footnote{current default for dimmest star is $V_{mag} \leq 5$} 
in the current filter band, there would be no sources within 12 arc-seconds having a brightness of $V_{mag}+3 $ or
brighter.

The EXCAM star acquisition algorithm uses maxima search, centroiding, and energy content
of the neighborhood around the peak pixel,  to find the star in the image.  Once a star has been
identified within the image, the camera integrates a photon signal to a high enough SNR. The
centroiding algorithm computes a star location within the region of interest in terms of a pixel
column and row.

For the first EXCAM star acquisition, the star's location is expected to be as far as four arc seconds from the EXCAM target pixel location. 
For a successful star acquisition, the star must be
placed at a central pixel location around the target within the EXCAM capture range.   The capture range has been calculated to be 0.054 arc-seconds radius considering ACS  errors, high-frequency jitter, and centroiding errors that limit the star's placement at a desired target pixel location.   If the star's location falls outside the desired pixel threshold,  CGI will employ the Roman ACS to adjust the pointing.  This will involve exchanging information between the CGI and the Roman ACS using CGI Flight Software (\textit{iFSW}) to get the star within that pixel threshold.  Therefore, the EXCAM star acquisition algorithm calculates the star's centroid location and its relative position with respect to the desired pixel.  This value would be offloaded to the spacecraft ACS for correction.   
After the S/C ACS informs the CGI that correction has been made, the instrument takes a new set of EXCAM frames and restarts the acquisition process, computes the new star location's centroid, and repeats sending offloads to the S/C ACS for correction until the star is placed within the desired threshold; at this point, EXCAM star acquisition has successfully acquired a star.  This threshold value is a settable parameter and represents the EXCAM star acquisition capture range as noted above.  

For coronagraphic observations, the ability of the EXCAM star acquisition to place the star within this threshold puts the star within a radius that is within the capture range of the Low Order
 Wavefront Sensing and Control (LOWFSC) system\cite{BD2022,BK2024}; the next step toward the ultimate objective of aligning the star on the FPM after the star has been acquired.  Once the FPM comes into the optical path, the star’s light will reflect onto the LOWFSC detector, LOCAM, and LOWFSC will generate tip/tilt measurements to be able to close the LoS loop\cite{MM2024,JS2024} and allow for the loop to stabilize.  At this point, the star must be centered, and the LoS loop would be closed and sending offloads, which would be the FSM quasi-static drift measured by piezoelectric actuators (PZT) strain gauges to S/C ACS for correction. Throughout this process, CGI tracking states and ACS status flags are being communicated between the CGI and the spacecraft, indicating the different stages of star acquisition.

During the CGI TVAC testing campaign, both bright and dim stars were placed within and beyond a 4-arcsecond radius from the target pixel, and EXCAM star acquisition algorithms were tested. The results will be presented in the following sections.

\subsection{Raster Scan Star Acquisition Mode}
\label{sec:raster}
\begin{figure}
    \centering
    \includegraphics[width=1.0\linewidth]{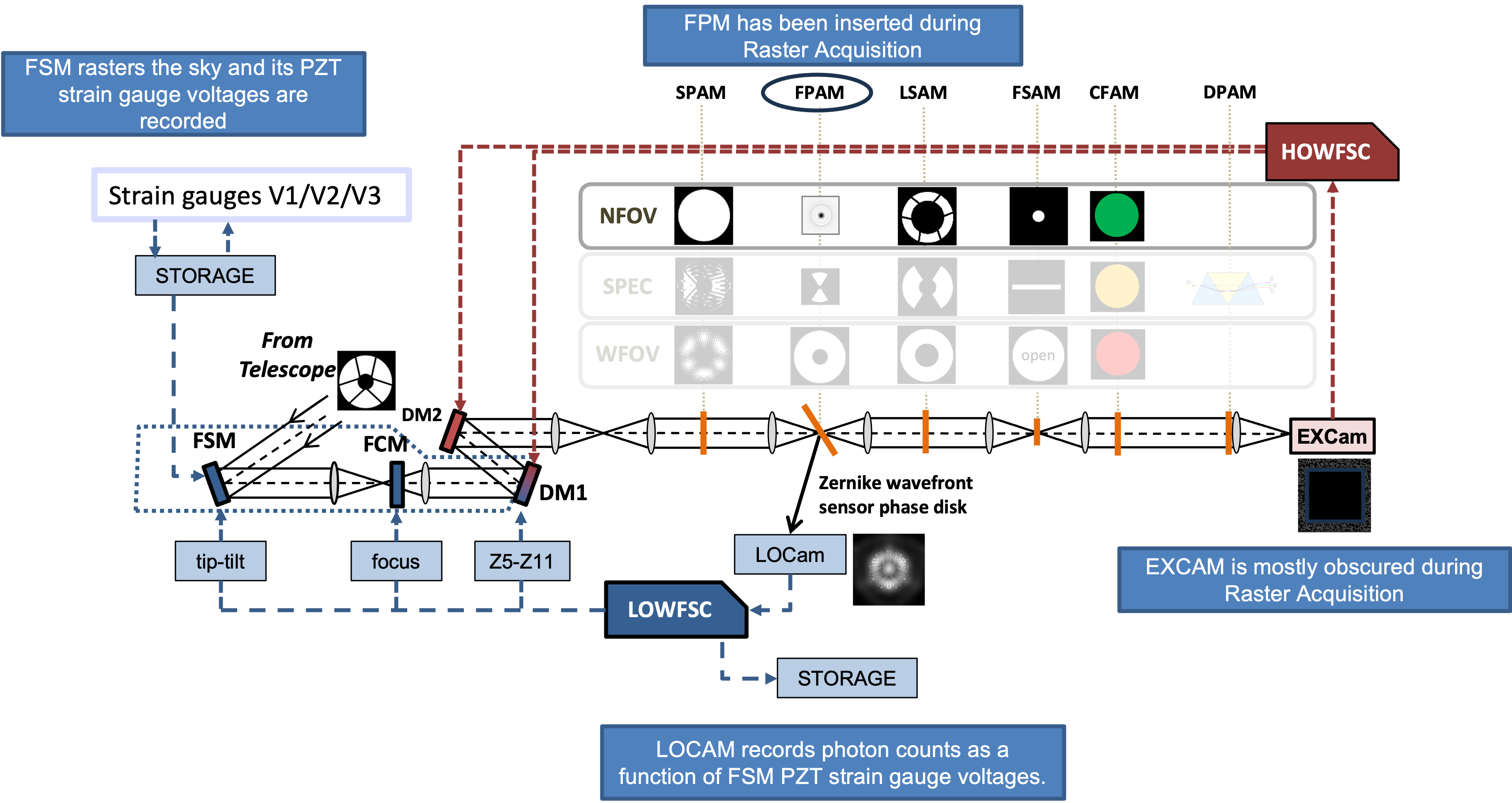}
     \caption{The CGI optical configuration during raster scan star acquisition when the FPM has been inserted
     into the optical path. The EXCAM is obscured, and the CGI uses the FSM and LOCAM to acquire
    a star. The ACS will place the star within a 0.3" radius of the detector center.  The FSM will traverse
     its full range of motion. The scanning motion of the FSM will correspond to a pre-recorded voltage
     profile for each of FSM's three PZTs strain gauges.  FSM profiles are stored in MRAM, with each
     profile describing a different motion pattern for the three PZTs. A photon count will be recorded at 
     1000 Hz as a function of the FSM PZT strain-gauge voltages implemented in the LOWFSC FPGA.
     The raster scan star acquisition  algorithm uses the raster image to find the star in the image.
     The ACS will bring the star near the center of the FPM.  The LoS loop will be closed when a raster
  scan star acquisition is successful.}
    \label{fig:lowfs_raster}
\end{figure}

Raster Scan Star Acquisition is performed after the reference star has been acquired through EXCAM star acquisition. All calibrations to set up the dark hole must also be completed. Finally, the required image contrast must be achieved. At this point, the CGI FPM is on the optical path. 
Moving the Precision Alignment Mechanisms (PAMs) could alter the quality of the DH contrast.
As shown in Fig.~\ref{fig:lowfs_raster}, there are six PAMs on the CGI optical path: Shape Pupil Alignment
Mechanism (SPAM), Focal Plane Alignment Mechanism (FPAM), Lyot Stop Alignment 
 Mechanism (LSAM), Field Stop Alignment Mechanism (FSAM), Color Filter Alignment
 Mechanism (CFAM), and Dispersion/Polarization Alignment Mechanism (DPAM). The main
purpose of the PAM assembly is to move the selected PAM to a commanded position so that the 
correct optical element is in the beam path within the required precision.
Since the CGI FPM is on the optical path, it blocks most of the EXCAM field of view, and the EXCAM star acquisition method is not feasible.  The FSM 
 and the pupil-plane LOWFS detector, LOCAM are utilized in concert to perform this secondary raster acquisition.

Fig.~\ref{fig:lowfs_raster} illustrates the optical configuration during the Raster star acquisition after the FPM has been inserted into the optical path and the EXCAM detector is obscured.  At this point, the required image contrast has been achieved, and moving the PAMs could alter the quality of the DH contrast.

Raster Scan Star Acquisition starts after each slew and roll during CGI observations.   These observations are defined in the CGI Observing Scenarios (OS) as shown in Fig.~\ref{fig:os11}.  Each observation cycle includes acquiring a bright reference star and then slew to a dim target and then rolling about the same target star to reacquire and repeat.  There are around 10-star acquisition operations per each observation cycle.  For more information on CGI observing scenarios refer to this paper\cite{JK2024} in the reference.

Like the EXCAM star acquisition, the S/C ACS would provide extra pointing capability during the raster scan star acquisition.
During CGI technology demonstration observations, the spacecraft's attitude control system (ACS) achieves better-pointing accuracy than during the initial star acquisition using the EXCAM method, allowing the ACS to place the star within 0.3 arc-seconds of the detector center.  This is an interface requirement between the CGI and the ACS systems to guarantee successful CGI acquisition and reacquisition during slews and rolls.  ACS pointing is now limited by CGI and WFI boresight drift due to thermal disturbances, ACS  errors, and high-frequency jitter.

The raster acquisition method includes scanning the sky with the FSM and recording light-intensity counts using LOCAM. The FSM will traverse its full range of motion, and a photon count will be recorded at 1000 Hz as a function of the FSM PZT strain-gauge voltages, implemented in Low Order Wave Front Sensing and Control (LOWFSC) field programmable gate array (FPGA).
 The scanning motion of the FSM will correspond to a pre-recorded voltage profile for each of FSM's three  PZTs strain gauges, whose measurement is triggered externally by the LOCAM.

Some setup is required before starting the raster scan star acquisition.    Depending on the target star of interest, the LOWFSC estimator will build a new estimator for the specified target star and its detector.  LOCAM gain needs to be set according to the star’s visual magnitude.  Additionally, to ensure the background is properly removed, a LOWFSC dark frame is specified for the LOCAM gain to be used with the star where the reference data is collected.  A specific raster trajectory must be loaded for each mask before initiating raster scan star acquisition. This depends on the coronagraphic mask that is being used.  The star's point-spread function (PSF) becomes highly dependent on the CGI mode of operation: 
Hybrid Lyot Coronagraph(HLC), Shaped Pupil Coronagraph(SPC),
and SPC Spectroscopy (SPEC).

\begin{figure}
    \centering
    \includegraphics[width=0.65\linewidth]{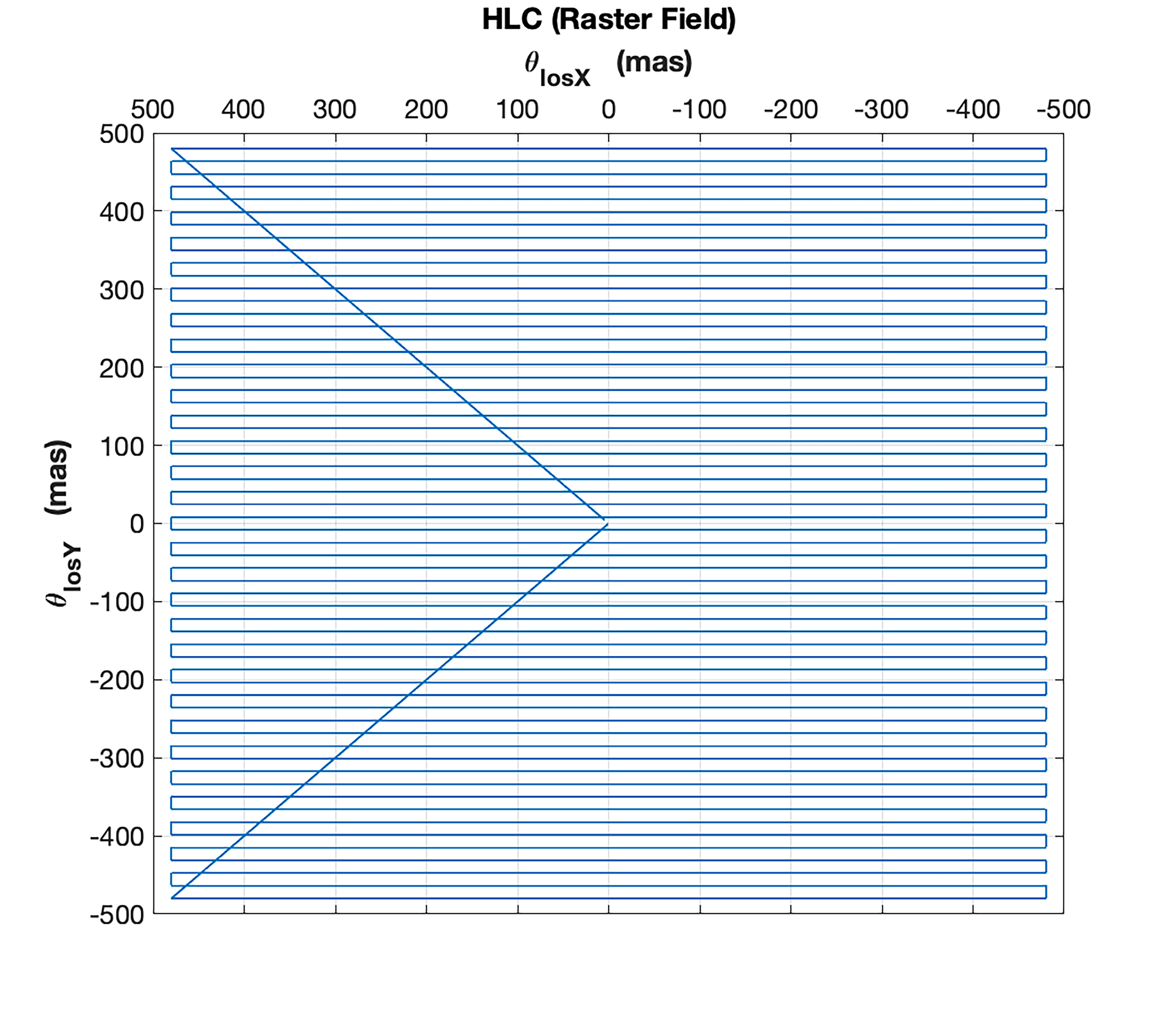}
    \caption{Modeled FSM raster trajectory for the HLC raster field of  $0.48" \times 0.48"$, on the sky, with
    initial movements to the corner of the square and a return to the center at the end of the raster.}    \label{fig:HLC_raster}
\end{figure}

Fig.~\ref{fig:HLC_raster} is an example of a raster scan trajectory for an HLC narrow field of view (NFOV) mask.  
This pattern is used to scan a star across a focal-plane mask, in this case, the circular HLC mask for Band 1 NFOV operation, to determine an FSM placement that centers the star on the mask.  It assumes Roman has pointed to a star within $0.3$ arc-seconds of the detector center.  HLC raster moves the FSM in a square-shaped raster, an area of $0.48 \times 0.48$ arc-seconds on the sky, with initial movements to the corner of the square and a return to the center at the end of the raster.  This profile contains $60 \times 60$ grid points in both columns and rows as the FSM rasters across the entire range.  The pixels in the image are 16 milli-arc-seconds (mas on the sky) away from each other.   Depending on feature size, the raster grid size and shape are different for each mode.

\begin{figure}
    \centering
    \includegraphics[width=1.0\linewidth]{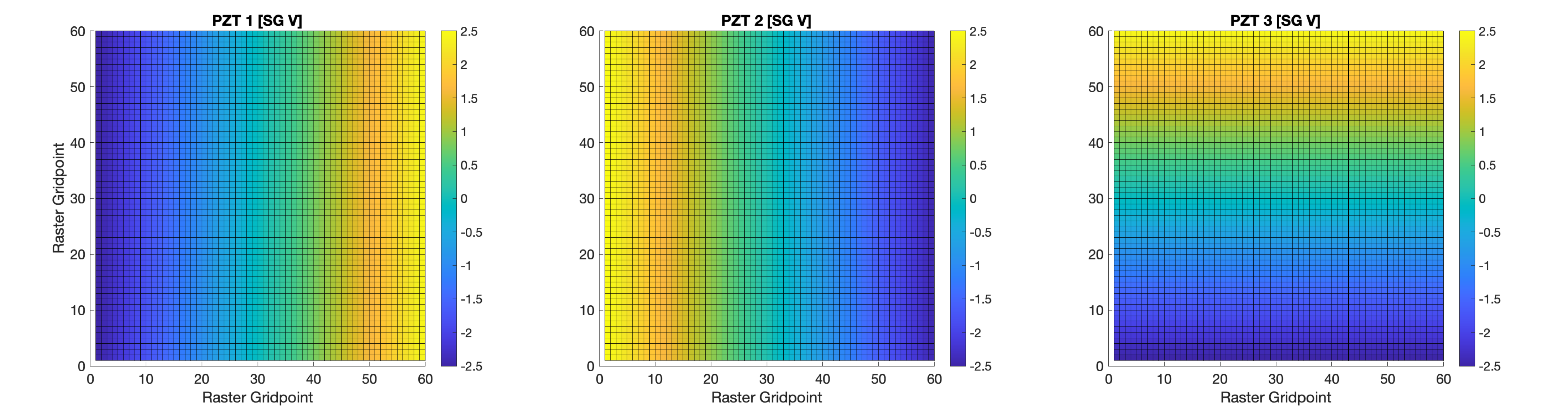}
    \caption{A conceptual example of a map of collected FSM strain gauge (SG) voltages for all 3 PZTs across
     an HLC raster grid size of $ 60  \times 60 $ grid points.}
    \label{fig:HLC_SGmaps}
\end{figure}

To identify the star's location within the image, \textit{iFSW} prepares a raster image. Each of the three PZTs creates a strain gauge map by reading strain gauge voltages across the raster image. 
Fig.~\ref{fig:HLC_SGmaps}   shows a conceptual example of a map of collected FSM strain gauge voltages for all three PZTs across a raster image with $ 60  \times 60 $ raster grid points.

If the light is reflected from the FPM to LOCAM, the light will be registered at that specific location. If the light does not bounce off the FPM, the LOCAM will register no photon counts except noise. The photon counts collected at each new position in the sky will then be put into a matrix, effectively replicating a focal plane image, where the ``pixel plate scale” is equivalent to the distance between two consecutive LOCAM images.     The Raster Star ID algorithm uses the raster image to determine, through photometry, whether the scan data shows the star in the image.

To determine if the current pixel is part of the feature, its total value must be less than a threshold value.  For HLC NFOV and SPC wide field of view(WFOV) masks, the feature is a bright spot in the middle and has more light since the light is reflected in LOCAM.    However, for SPEC masks, where the feature is in the lobes of the bowtie, it has less light because light passes through to EXCAM.  
 If the star is NOT in the image, the \textit{iFSW} issues an error message and exits the star acquisition activity. 

\begin{figure}
    \centering
    \includegraphics[width=0.85\linewidth]{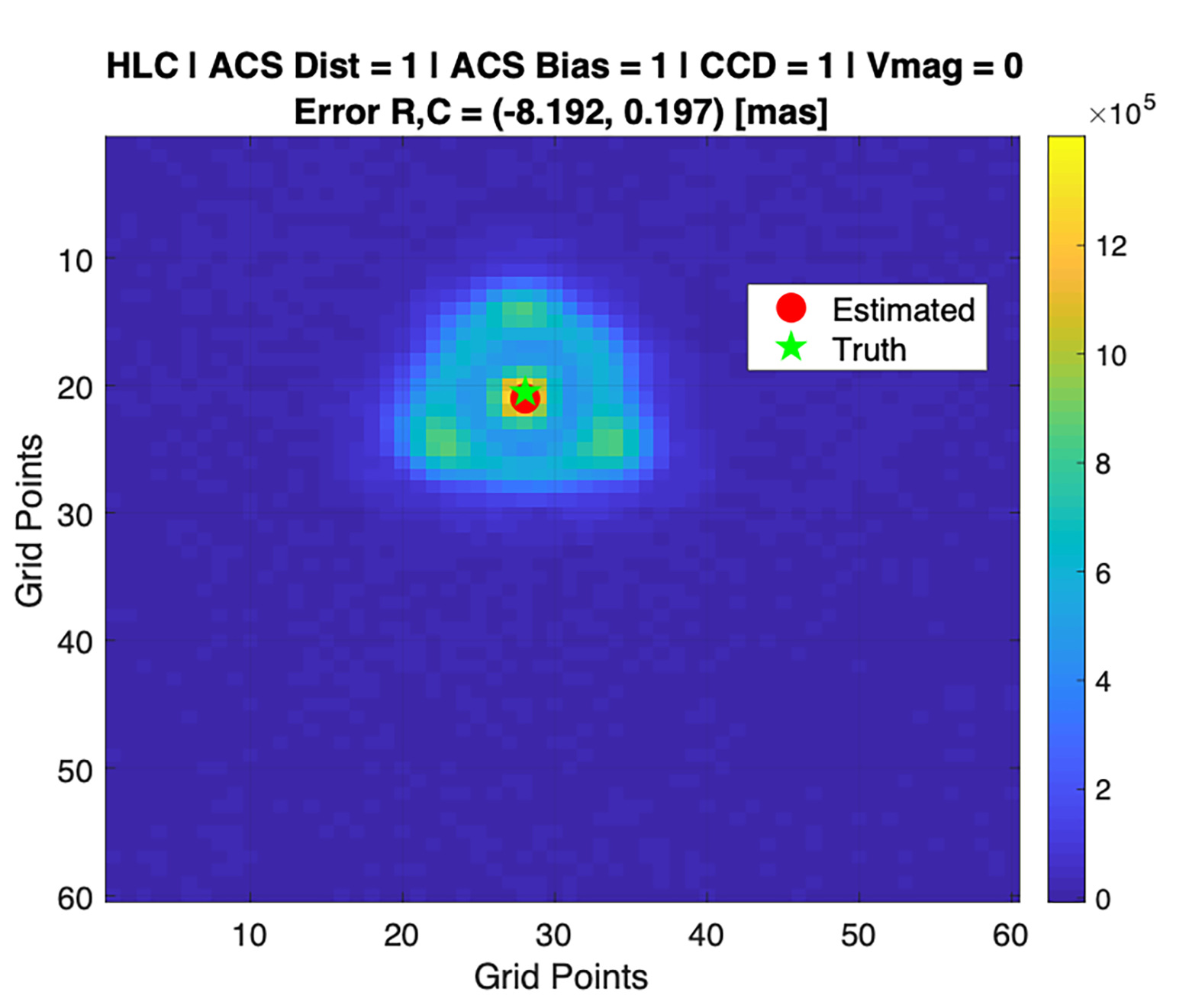}
        \caption{An example of a star location computed by the \textit{iFSW} in an HLC raster image of $60 \times 60$ grid
   points as a function of photon counts, and then compared with its true location.}
    \label{fig:HLC_image}
\end{figure}

If raster Star ID is successful, \textit{iFSW} calls the Raster Find Star algorithm, which determines the grid-point location of the star where LOWFSC FPGA collects LOCAM photon counts across the raster image for an HLC  mask.  The algorithm performs a bilinear interpolation, or single-pixel evaluation, depending on how far the calculated center of the mask is from the center of a pixel and outputs PZT voltage values corresponding to that raster grid point. 
Then the  \textit{iFSW} assembles LOCAM counts image and strain gauge maps and computes centroid in LOCAM image.  Fig.~\ref{fig:HLC_image} shows an example of a star's location in a raster image,  as computed by the \textit{iFSW} and then compared with its true location.

 This pixel location, represented in terms of PZT voltage values, would be mapped to the corresponding delta-H and delta-V and then offloaded to the ACS to bring the target star near the center of the mask.
 Offloads are defined as the FSM strain gauge voltages required for the PZTs to reach a desired grid point location during the raster scan star acquisition. These voltages would be mapped to delta-H and delta-V in radians. Section~\ref{sec:offloads} defines offloads in more detail.
 After the correction,  the star's center falls at least partially on the wavefront sensor dimple, enabling phase measurements.  
 When the star is centered on the mask, the \textit{iFSW}  can close the LoS loop to complete a successful raster scan star acquisition.

To keep the FSM centered within its range, the FSM position, as measured by the strain gauges, will be gradually offloaded to the spacecraft ACS when the LoS loop is active.
The \textit{iFSW} then monitors the three FSM strain gauge voltages to determine if they are within their setpoint threshold.  
After the star is acquired, CGI actively tracks it in concert with the ACS, while the LOWFSC system maintains the dark hole and starlight suppression.

Examples of raster trajectories and raster scan star acquisition will be provided in the following sections.

\subsubsection{Raster Trajectories}
\label{sec:rastertrajectories}
One of the main capabilities of the FSM assembly is the ability to execute raster scan trajectories from the LOWFSC slice. For more information on FSM refer to this paper\cite{JS2021} in the reference.
Before executing the raster motion, the desired profile would be loaded into the LOWFSC FPGA, and then executed by it to move the FSM in a pre-defined manner at high speed. 
Each raster has prefilters associated with it to smooth the steps. The voltage slew rates associated with the residual step responses cause variations in the current. 

The \textit{iFSW} stores these FSM profiles in Magnetoresistive Random Access Memory (MRAM), with each profile describing a different motion pattern for the three PZTs of the FSM.
It is important to note that the FSM ``inner loop” that is implemented in the FPGA would be closed with strain gauge feedback about each of the three PZTs during the raster scan.  The sampling rate for this loop is at $10$ kHz.

Raster scan mask type specifies whether the raster signal is looking at the reflective light from a disk mask, which is used for HLC NFOV and SPC WFOV, or from a mask with a bowtie hole for SPEC.  CGI has a specific FSM raster trajectory for each mask: HLC NFOV, SPC WFOV,  
SPC SPEC with center wavelength $  \lambda = 730nm$ (SPC SPEC730), and SPC SPEC with center
wavelength $ \lambda = 660nm$ (SPC SPEC660).  In addition, there is a raster trajectory for the EXCAM detector calibration called Flat Fielding. These raster patterns are designed to scan across a specific mask, and each has different feature sizes.  The raster range will be larger for masks with larger features.  The smallest raster range is for the HLC NFOV mode and is $0.48 \times 0.48$ (arc-seconds, on the sky), and the largest raster scan range is for SPC SPEC730 mode and is $0.94 \times 0.66$ (arc-seconds, on the sky) that covers features of interest for that mask.  By default, each raster trajectory takes 160 seconds to complete. However, this duration can be updated by changing a parameter in \textit{iFSW}, specifically the ``wait and stare" parameter, which is the count of the number of  $100 \mu s $ steps the FPGA should wait at each point of the profile which is currently 16 milli-seconds(ms).  Default raster trajectories have the same duration for all modes, and for smaller range trajectories, zero-padding was added.  
It is important to note that a few memory slots are dedicated to SPARE, which will be used to store any future potential profile used to drive the PZTs in a more customized manner if needed.

To design these trajectories, the desired tip/tilt of the FSM mechanical frame was mapped to its Actuation Point Displacement (APD) for each of the FSM three PZTs since FSM motion is commanded by  \textit{iFSW} using APDs, and not PZT voltages. APD motion is negative of FSM PZTs motion.  When FSM PZTs extend the APD ``retracts” (negative APD motion) and moves the mirror back in the -z-axis direction of the FSM mechanical frame. Conversely, when FSM PZTs contract the APD extends outward (positive APD motion) or in the +z-axis of the FSM mechanical frame.  FSM PZT motion is monitored by strain gauges attached to them.  There is a steering matrix that is used internally by the FPGA that provides a mapping between the tip/tilt Zernikes (Z2 and Z3) and the PZT position as represented by APD.  

\begin{figure}
    \centering
    \includegraphics[width=0.85\linewidth]{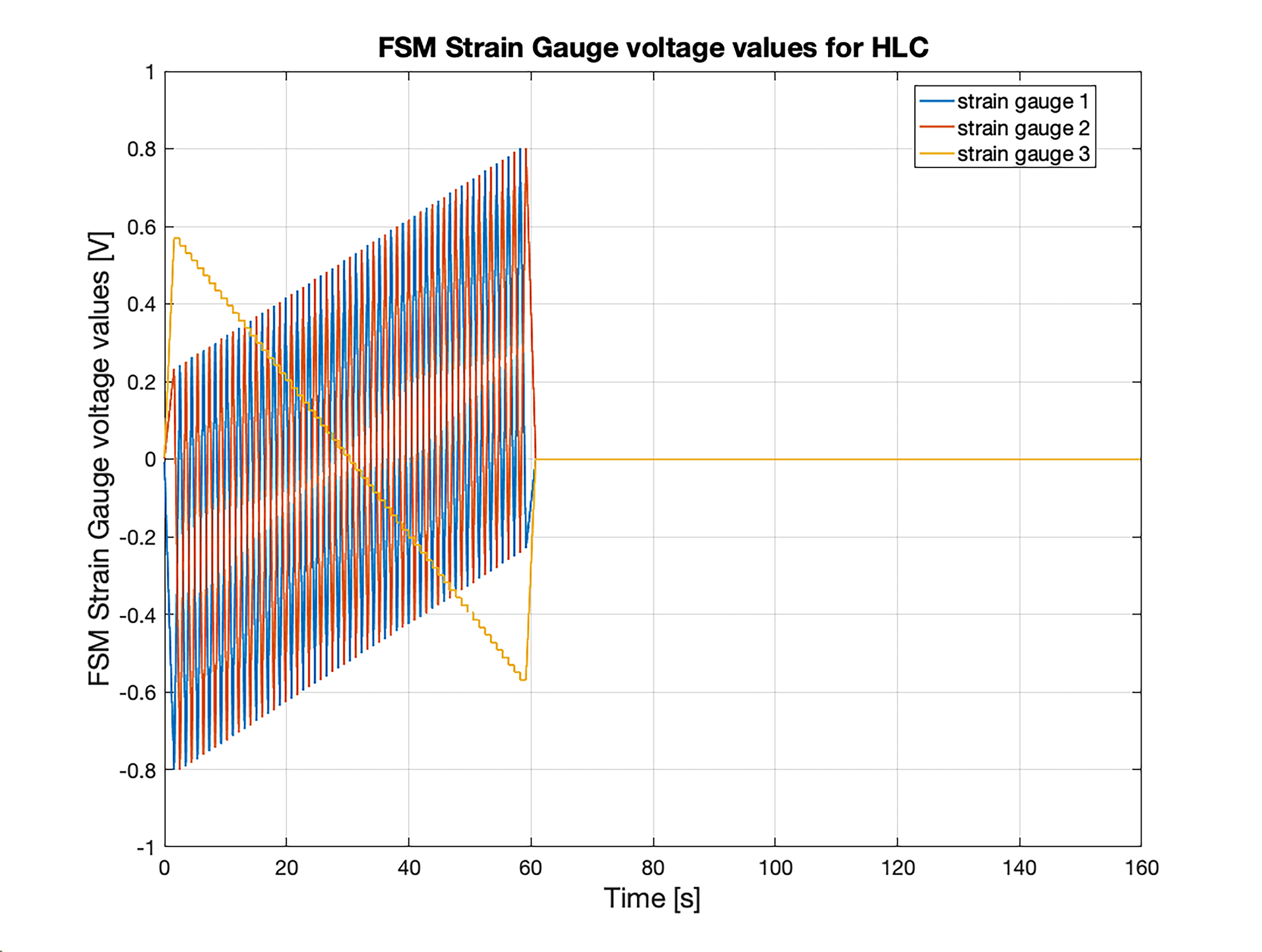}
    \caption{FSM PZTs three strain gauge (SG) voltage profiles for the HLC trajectory.  These profiles are stored
    in the MRAM, with each profile describing a different motion pattern for the three PZTs of the FSM.
    The trajectory completes in less than 160 seconds, and it is zero-padded for the remaining of the
     duration.}
    \label{fig:HLC_SG}
\end{figure}

The algorithm in \textit{iFSW} calculates the associated APD in microns based on the requested strain gauge voltage values for input into the FPGA. Fig.~\ref{fig:HLC_SG} shows FSM  strain gauge voltages for an HLC trajectory that completes in less than 160 seconds and as shown has been zero-padded for the rest of the duration. 
All designed trajectories were tested with the FSM and other hardware and results show that measured and designed trajectories match.

\subsection{CGI Tracking state}
\label{sec:tracking}

\begin{figure}
    \centering
    \includegraphics[width=1\linewidth]{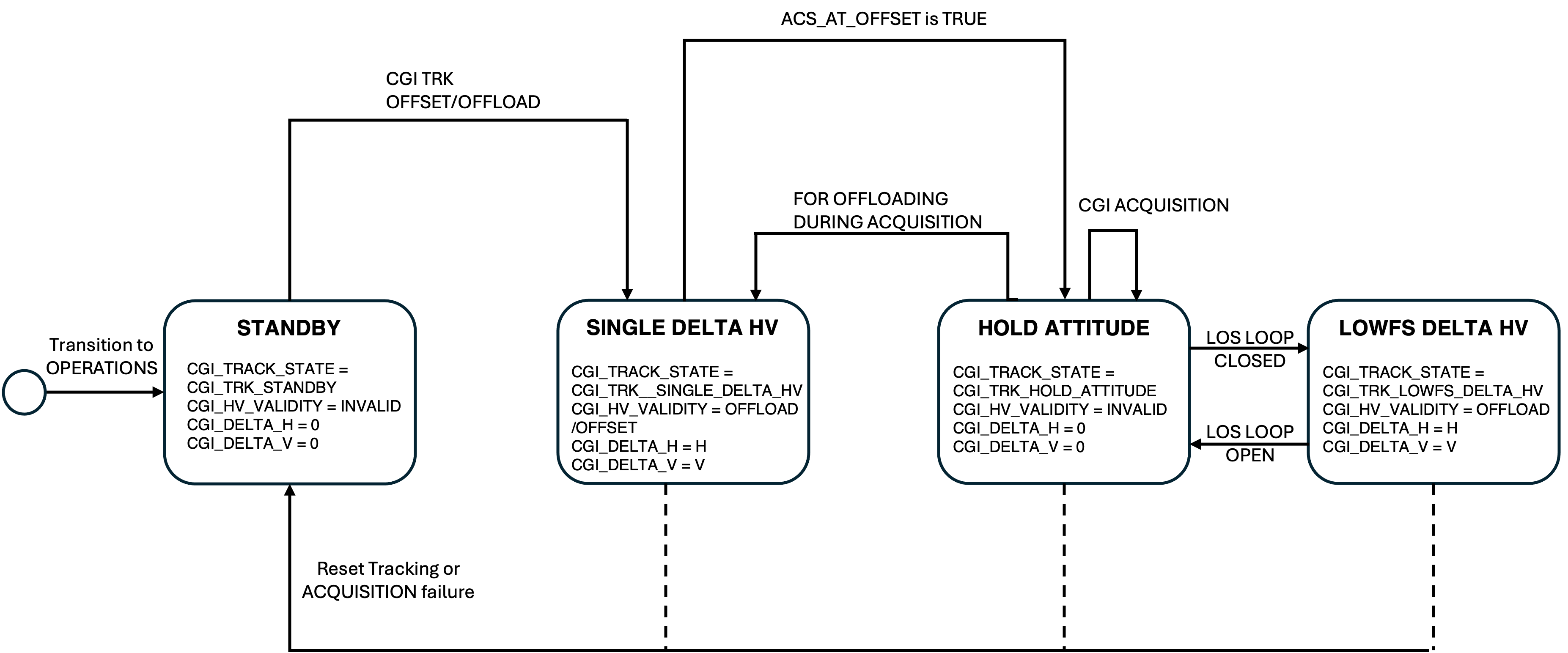}
    \caption{CGI Tracking State Diagram during the CGI Star Acquisition.  Each tracking state represents a setting to one of the CGI-defined states and provides the spacecraft ACS with information about how the spacecraft should react to it.}
    \label{fig:ACQ_TRK_STATE}
\end{figure}

During CGI star acquisition and observation, the RST ACS enhances the instrument's pointing capabilities, enabled through communication between the CGI and RST flight software.
During CGI operations, information is exchanged between the CGI and the ACS through incoming packets
rom the RST (at $4 Hz$ rate) and outgoing packets from the CGI (at $1 Hz$ rate). For this purpose, a state machine is used and the transitions in the state machine model represent the conditions under which CGI FSW should update the values of its outgoing packet as shown in a simplified block diagram Fig.~\ref{fig:ACQ_TRK_STATE}.  

A key field in the CGI outgoing packet is the tracking state during the star acquisition.   Each tracking state represents a setting to one of the CGI-defined states and provides the spacecraft ACS with information about how the spacecraft should react to it.  The initial tracking state is the ``standby” state when the instrument is transitioned to the operational mode.  After the instrument is prepared to perform various activities, such as star acquisition, the tracking state transitions to other states.   

For CGI star acquisition to start, the tracking state must be in ``hold attitude" where the S/C has reached its required pointing stability.  This would be communicated to the CGI by a flag from the ACS, indicating that S/C is stable enough and attitude errors are within tolerance levels.  
When S/C attitude errors are outside the tolerance, this same flag from S/C ACS is set to false.  
Nominally, it takes less than 5 minutes for the ACS to reach its required pointing stability for the
CGI observations.  More detailed discussions regarding ACS disturbance events are presented
in the Sec.~\ref{sec:ACSflags}.

Other important fields in the CGI packet are the ``offloads” and ``offsets". 
In either case, S/C ACS uses these offloads/offsets and treats them as attitude errors to bring the star to the desired location that the  CGI requests.   There would be two CGI tracking states: ``single delta hv'' and ``lowfs delta hv'' when S/C ACS expects these ``offloads” from the CGI.    During the CGI star acquisition, when the tracking state is in ``single delta hv'', the CGI requires pointing correction from the ACS system.  
When the tracking state is in ``lowfs delta hv", it means that star acquisition is completed, the LoS loop is closed, and offloads to ACS  are FSM strain gauge drifts that spacecraft would correct for to center the FSM and bring it to its home position.
At this point, the instrument is ready for tech demo observations.

 \subsubsection{CGI offloads/offsets to the spacecraft}
 \label{sec:offloads}
As was mentioned in the previous section, other important fields in the CGI outgoing packet are the ``offload'' and ``offset" that CGI sends to the S/C.  
The spacecraft ACS  treats offloads as attitude errors to bring the star to the desired location that CGI is requesting and can be accumulated until the spacecraft corrects for them.  
Offloads are considered to be relative motions.   Examples of CGI offloads are star location on EXCAM relative to a pixel (centroid location), or FSM strain gauge voltages for PZTs to reach a desired grid point location during raster scan star acquisition or FSM motion away from its desired home position. 

Offsets are desired motions relative to a pre-determined setpoint and are considered to be absolute motions.  For example, if we don’t find the star at the current location, CGI will send an offset to S/C ACS through \textit{iFSW} for a new location.   ACS implements these offsets as a new location and resets any prior offload accumulation before setting this new offset.  

\begin{figure}
    \centering
    \includegraphics[width=0.75\linewidth]{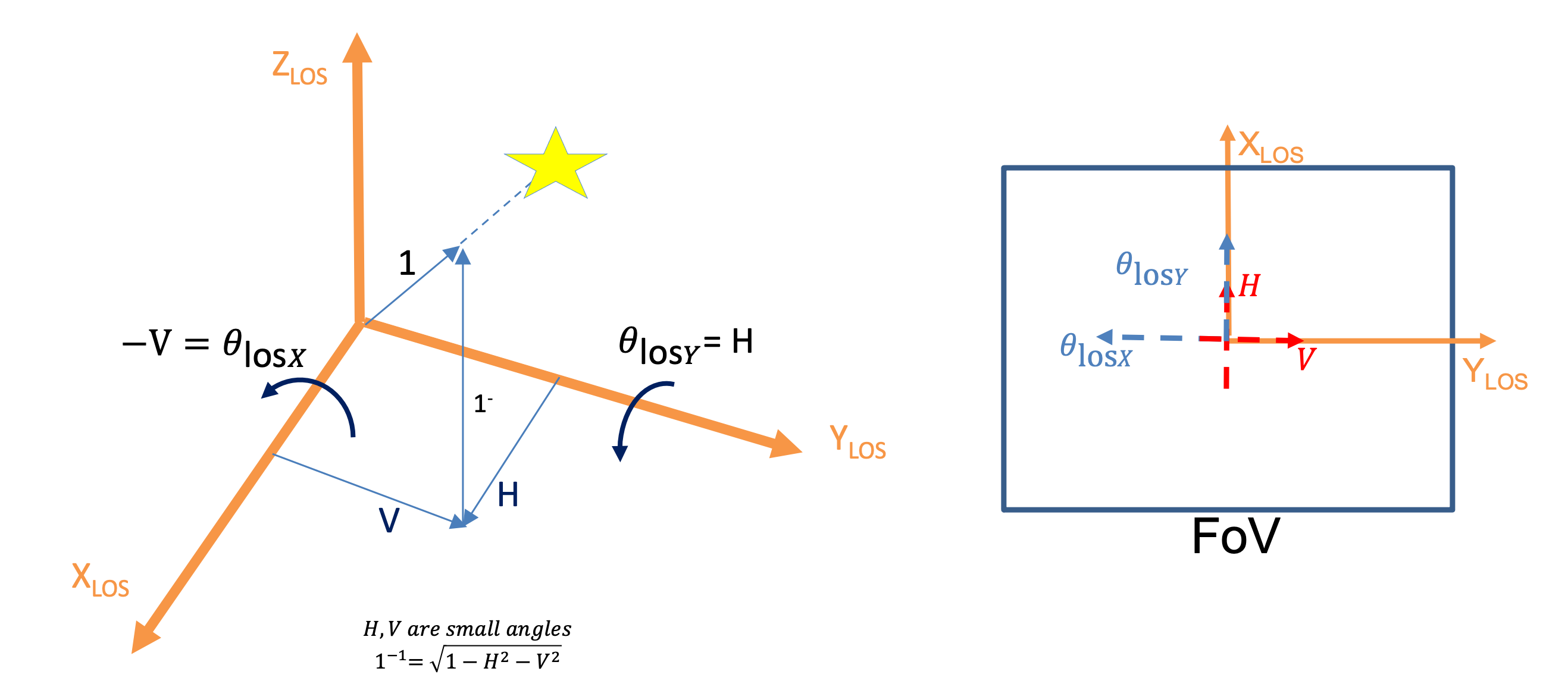}
    \caption{CGI offloads delta-H (or H) and delta-V (or V) to the S/C ACS in the CGI LoS frame: delta-H is
    the CGI target star direction vector along the CGI X axis ($X_{LoS}$), and delta-V is the CGI target
    star direction vectoralong CGI Y axis $(Y_{LoS})$.  Delta-H/V rotation is expected to align CGI 
     LoS Z-axis $(Z_{LoS})$ to the star vector. Upon receipt of offloads delta-H/V in radians,  ACS
     rotates the S/C along $X_{LoS}$ by $\theta_{LoSX}$ or -delta-V and along $Y_{LoS}$ by 
     $\theta_{LoSY}$  or delta-H.}
    \label{fig:LoS_frame}
\end{figure}

Each offload or offset has two components in a unit of radians:  delta-H, CGI target star direction vector along CGI X axis ($X_{LoS}$), and delta-V, CGI target star direction vector along CGI Y axis $(Y_{LoS})$. The CGI LoS frame has been defined and agreed upon between the CGI and the ACS  teams so the ACS would properly correct these offloads or offsets.  The relation between delta-H and delta-V (or equivalently H/V) and CGI LoS frame rotation angles is given in Fig.~\ref{fig:LoS_frame} for small angle rotations about $X_{LoS}, Y_{LoS}$ respectively. 

The resulting delta-H/V rotation is expected to align CGI LoS Z-axis $(Z_{LoS})$ to the star vector.  Upon receipt of offloads delta-H/V in radians,  ACS rotates the S/C along $X_{LoS}$ by $\theta_{LoSX}$ or -delta-V and $Y_{LoS}$ by $\theta_{LoSY}$  or delta-H as shown in Fig.~\ref{fig:LoS_frame}.

\subsubsection{ACS flags}
 \label{sec:ACSflags}

 During CGI operations, RST communicates the ACS status to the CGI through flags transmitted via FSW communication. These flags indicate whether the spacecraft maintains the necessary pointing stability throughout CGI operations.  
 
 One such flag, “AT Offset,” signals if the ACS attitude errors remain within acceptable threshold limits during CGI observations. If this condition occurs then the CGI tracking state transitions to “hold attitude,” thus allowing for the commencement of star acquisition. The flag will switch to FALSE if the ACS errors exceed the prescribed limits, indicating that the ACS must execute rotations to correct for any attitude errors such as CGI offloads.
  
 Additionally, two other flags may be triggered at any time, indicating larger-than-expected observatory motion resulting from reaction wheel (RW) zero-crossing or high gain antenna (HGA) movements, and causing large ACS errors.  
 This necessitates a pause in CGI observations until the ACS achieves its required pointing stability.  
 These ACS disturbance flags are called “RW Zero Crossing” and “HGA Move."   These events happen during the CGI observing scenario, as shown in Fig.~\ref{fig:os11}, where ACS sets the disturbance flags and pauses the CGI observations.
 When any of these flags are set to true, the CGI FSW must take action to protect the cameras by either turning them off or placing them in an IDLE mode while opening any closed control loops. Once these large observatory motions are completed, the ACS will reset these flags to false, prompting CGI to resume operations.  Nominally, it takes less than 5 minutes for the ACS to reach its required pointing stability and to resume nominal operation.
 
For a typical tech demo observing scenario, we should expect around 20-30 of these ACS disturbance events over several days.
During CGI star acquisition, TVAC tests, various ground commands, including both modes of star acquisition, were tested while ACS Disturbance flags were set to true to demonstrate that  \textit{iFSW} responds to these flags properly and recovers from them successfully.  These results are not shown in this paper.

\section{CGI Star Acquisition Testing at JPL}
\label{sec:testing}

Testing star acquisition and control algorithms at the CGI system level has always been a challenging task.  
During earlier development of these algorithms, a lot of modeling and simulation was performed to verify performance using analysis\cite{BD2022,JS2024}.
During the earlier stages, most of the design verification was conducted using models. Later, after implementing the algorithms in the flight software or FPGA, they were tested on various testbeds across multiple phases.
Part of the challenge lies in the interface between the spacecraft ACS and the CGI during star acquisition and Line-of-Sight (LoS) control, which requires collaboration and relevant interface requirements between the two systems.

Earlier in the Verification and Validation (V\&V) phase, the Functional Testbed (FTB) was used with engineering models and \textit{iFSW} for testing with limited functionality verification. As flight optics tests progressed, the CGI bench eventually got fully populated, and system-level testing of the flight article became possible.   During Full Functional Tests (FFT) CGI was tested with \textit{iFSW} at the system level in the air to verify the functionality of CGI system.  The last campaign was a repeat of FFT but in the thermal vacuum chamber (TVAC).  This is when CGI performance requirements were verified.  

CGI star acquisition testing requires a system-level testing platform and testing of its algorithms implemented in \textit{iFSW}  was done during multiple phases. 
CGI  Acquisition testing had multiple components: verification of star acquisition using the EXCAM, star acquisition using LOCAM and FSM for raster scan, testing of tracking state transitions during acquisition when the spacecraft and the CGI exchange information and finally testing of various commands where incoming ACS disturbance flags were turned ON and OFF to test proper flight software response to those flags.  These tests were done before the TVAC phase, during the FTB and FFT phases, which proved crucial for troubleshooting of \textit{iFSW}.

Both the CGI FFT and TVAC tests required extensive planning, with these campaigns being driven by cost and schedule constraints.  
Testing at earlier stages was helpful in troubleshooting star acquisition algorithms and led to successful star acquisition testing during the TVAC campaign.  

During TVAC testing we tested star acquisition algorithms with a fully populated flight CGI bench at the system level in a vacuum. Coronagraph Verification Stimulus (CVS) \cite{CN2024} was used to emulate a star as an input from the telescope while observing a source in space. It was also used to inject modeled flight-like perturbations and corrected when errors were ``offloaded” to it, acting as the spacecraft ACS during star acquisition tests.  

During the TVAC testing campaign, all star acquisition tests were performed for HLC NFOV
mode and there was no time allocated to test star acquisition for other modes.  
The following
reference\cite{GB2024} presents more details on the TVAC test configuration.
In test cases where disturbances
were enabled, the CVS injected modeled wavefront drift and LoS disturbances.  For more
information on these disturbances, see this paper\cite{MM2024} in the reference. 
 
\begin{figure}
    \centering
    \includegraphics[width=1.0\linewidth]{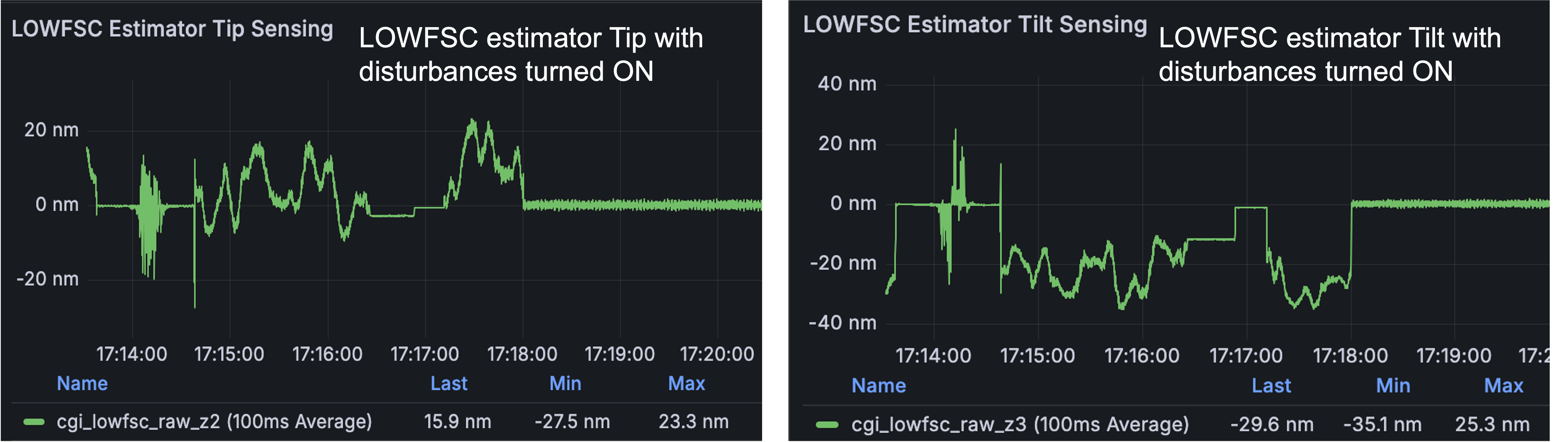}
    \caption{The CVS injected the required wavefront drift and LoS disturbances. LOWFSC estimator sensed the
    Z2/Z3 errors due to these disturbances.}
    \label{fig:Z2Z3_distOn}
\end{figure}

\begin{figure}
    \centering
    \includegraphics[width=1.0\linewidth]{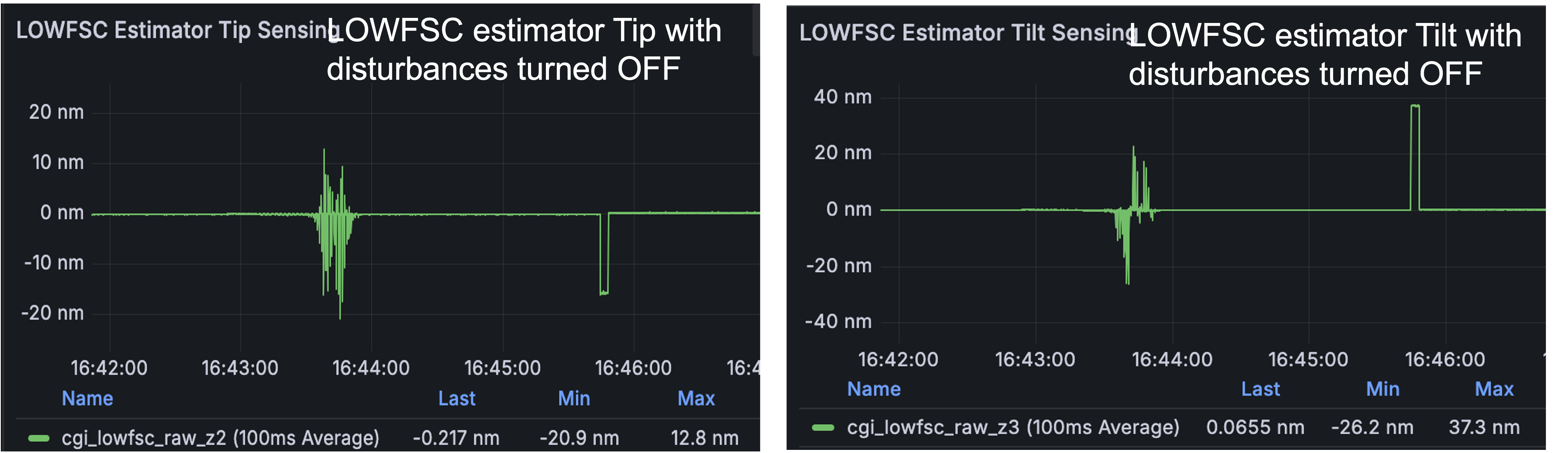}
    \caption{LOWFSC estimator sensed the Z2/Z3 errors for test cases where ACS and LoS disturbances were not
    turned ON. There was no jitter signature on Z2/Z3.}
    \label{fig:Z2Z3_distOff}
\end{figure}

 Figure~\ref{fig:Z2Z3_distOn} and Fig.~\ref{fig:Z2Z3_distOff} show two cases where ACS and LoS disturbances were turned on and off
 during raster scan star acquisition.  The LOWFSC estimator senses Z2/Z3 (tip/tilt) errors in both 
 cases and the jitter signature is shown for the case where LoS disturbances are turned on.

Telemetry data in the Grafana dashboard was used to visualize data during star acquisition tests and post-processing.

\subsection{EXCAM Star Acquisition tests in TVAC}
 \label{sec:excamtests}
Before testing of star acquisition in TVAC, several alignment and calibration tests of the CGI were conducted, which were essential for successful star acquisition. For more information, refer to these papers in reference\cite{AR2024,BK2024}.

To start a star acquisition in EXCAM mode, there were some steps to set up for the test, including PAMs setup.   EXCAM exposure time and gains were set based on the star’s visual magnitude, and its location was set up through the CVS.   We assumed that S/C ACS would place the star within a 4.0 arc-second radius of the CGI EXCAM center pixel.  During EXCAM star acquisition, the detector captures images, with the image locations represented in pixels on the detector.  Pixel notation is expressed as a [column, row] index.

During the TVAC tests, additional test cases were conducted where both dim and bright stars were placed at various locations within the detector's FoV, with disturbances enabled. For all these cases, we verified compliance with all EXCAM star acquisition requirements.  

The objective of these tests was to test algorithms in \textit{iFSW} and verify that when the star is placed at the required distance from the target pixel, the star acquisition algorithm can find the star and its location and computes the centroid and offset from desired pixel location to send as offload to the CVS for correction. 
  
 When the star was placed at the desired threshold region, we received a success message indicating that the EXCAM star acquisition was completed.  
We verified all related requirements,
and all test cases were successful. EXCAM star acquisition does not require a specific duration
for the entire process.  However, there are a couple of requirements for EXCAM StarID and
FindStar algorithms to be completed within 60 seconds and 20 seconds, respectively.
Both of those requirements were verified.
The next few examples show a few test cases demonstrating the CGI EXCAM star acquisition in TVAC.

\subsubsection{EXCAM Star Acquisition Example 1: Bright star at 2.24 arc-seconds radius from EXCAM Target pixel}
 \label{sec:excamex1}
\begin{figure}
    \centering
    \includegraphics[width=0.75\linewidth]{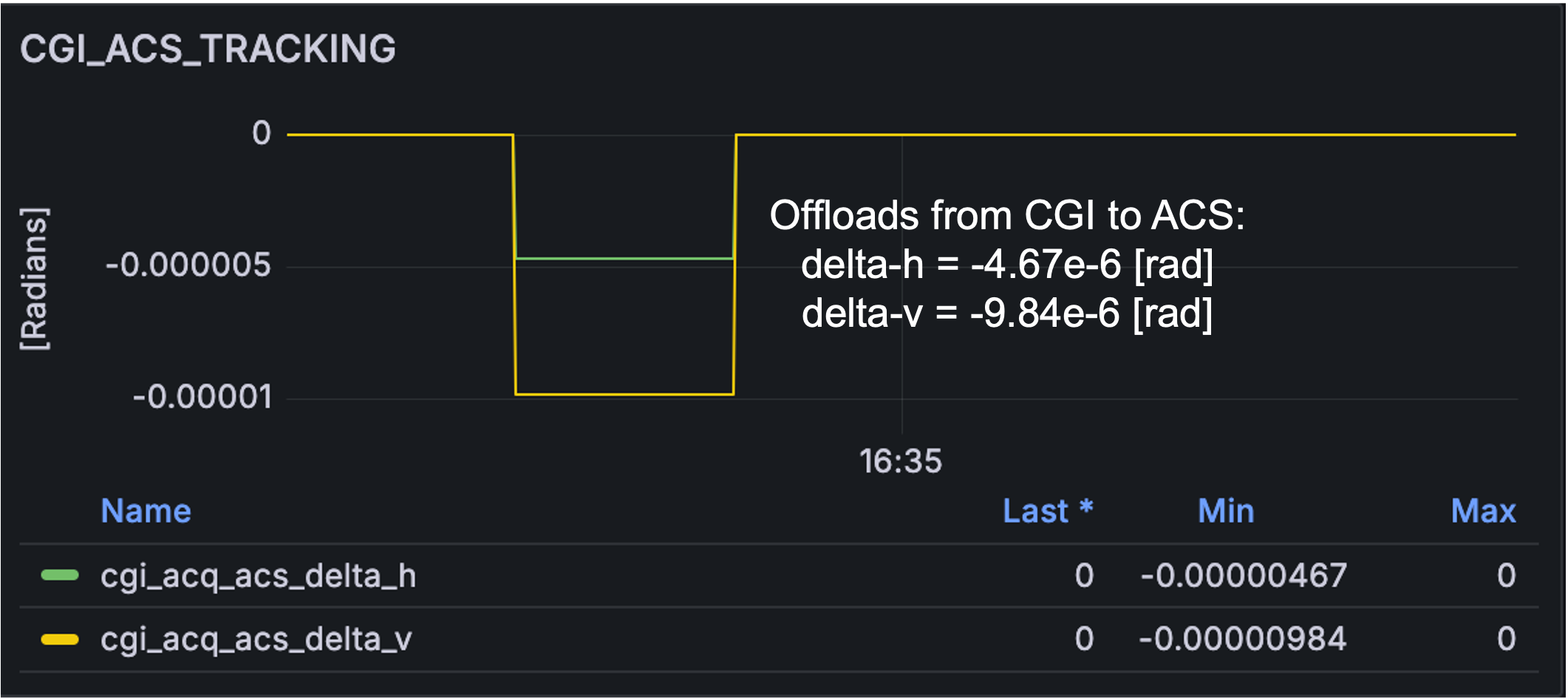}
    \caption{EXCAM Star Acquisition Offloads example 1:  The acquisition algorithm calculated the centroid,
    relative to the desired pixel location, mapped it to delta-H/V in radians, and then sent it to the CVS
    for correction. Delta-H/V corresponds to the location of PSF in the EXCAM.}
    \label{fig:EXCAM_ex1_deltaHV}
\end{figure}

\begin{figure}
    \centering
    \includegraphics[width=0.5\linewidth]{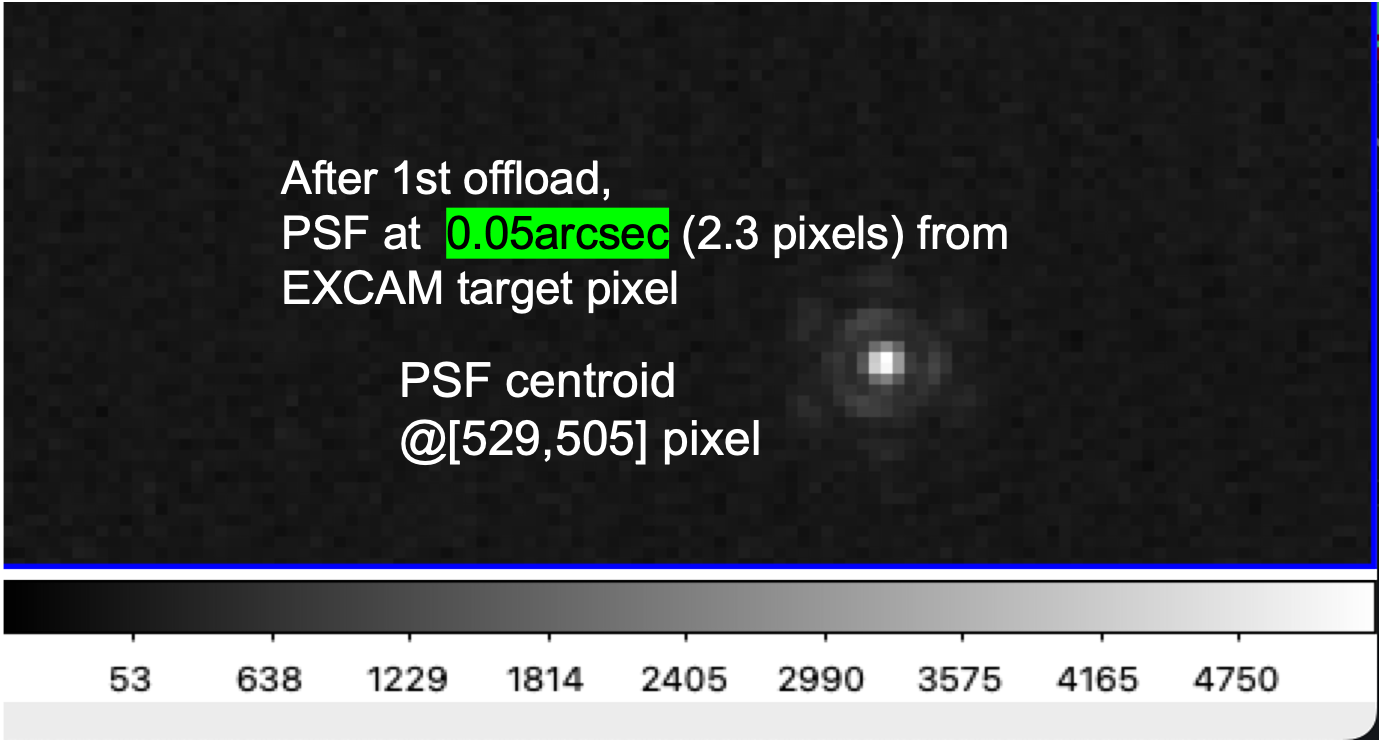}
    \caption{EXCAM star acquisition example 1: CVS adjusted pointing through an offload and placed the PSF
     within the EXCAM capture range of 2.5 pixels. The calculated centroid was within the desired
      pixel threshold.}
    \label{fig:EXCAM_fitsImage_example1}
\end{figure}

In this example, using the CVS, a star of visual magnitude 0 was placed at a 2.24 arc-seconds radius from the EXCAM target pixel and flight-like disturbances were turned on.  For this example, a neutral density filter, ND475 was used and the EXCAM gain was set to 1, the exposure time was set to 5 seconds, and the number of frames was set to 5.  The master dark was updated to match the commanded gain and exposure time.  Then the CGI tracking state was set to ``hold attitude” and the ground command for EXCAM star acquisition was sent through \textit{iFSW}.  After some initialization, EXCAM raw frames were cleaned from noise, and through the star acquisition algorithm star location was found to be at 2.242 arc-seconds from the target pixel and its centroid relative to the desired pixel location was computed as [Column, Row] in pixels. At this point, we verified that star acquisition algorithms have detected the star and its location correctly.   Then this value was mapped to delta-H/V in radians to be sent as offload to CVS for correction as shown in Fig.~\ref{fig:EXCAM_ex1_deltaHV} which shows telemetry from \textit{iFSW} in Grafana dashboard. During offloads, the tracking state would be in ``single delta hv”. The CGI sends these offloads to the ACS at a $1 Hz$ rate.
At this point, we could verify that the star's location relative to the target pixel was mapped to delta-H/V as expected. 
After the correction was made, the tracking state transitioned to ``hold attitude,” and the algorithm started over until the star was placed at the desired threshold location.  As shown in Fig.~\ref{fig:EXCAM_fitsImage_example1}, the new location of the star was computed to be 0.05 arc-seconds from the target pixel location within the desired threshold, and the EXCAM capture range to complete star acquisition successfully.  The star acquisition process for this example took around 4 minutes.

Post-processing of the data showed that star centroids matched expectations, and the star was placed at the desired threshold location within the EXCAM capture range.  This is to insert the FPM and align the star on the FPM to take an initial image and complete the initial acquisition process.

\subsubsection{EXCAM Star Acquisition Example 2: Bright star at 4.24 arc-seconds radius from EXCAM Target pixel}
 \label{sec:excamex2}
\begin{figure}
    \centering
    \includegraphics[width=1\linewidth]{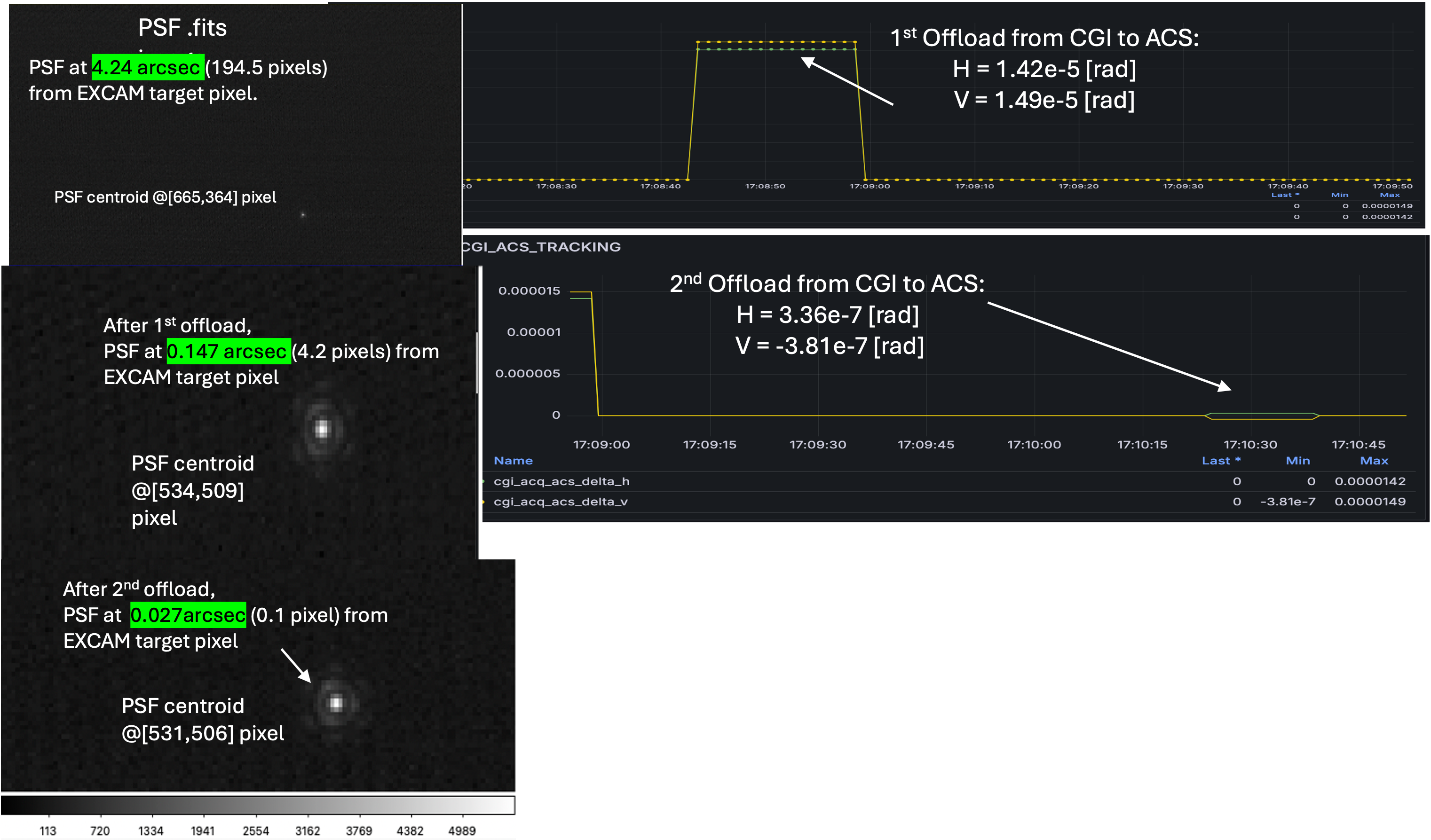}
    \caption{EXCAM star acquisition example 2: The star was placed outside the required  EXCAM FoV at 4.24"
    radius. Successful EXCAM star acquisition after 2 offloads and correction by the CVS.  The first
     offload and correction placed the star within 0.147" of the desired pixel location.  The second 
     offload/correction placed the star within 0.027" of the desired pixel location.  The process
      took around 5 minutes.}
    \label{fig:EXCAM_ex2_deltaHV}
\end{figure}

In this example, a star of visual magnitude 0 was placed at a $4.24$ arc-seconds radius from the target pixel to verify that we can still acquire a star even if it is slightly outside the required star acquisition camera’s FoV.  An interface requirement between the CGI and ACS requires the ACS to place the star within a $4"$ radius from the EXCAM target pixel. For this example, the same setup as example 1 was used.    In this case, it took two offloads until the psf was placed at 0.027 arc-seconds, well within the desired threshold as shown in Fig.~\ref{fig:EXCAM_ex2_deltaHV}.  The EXCAM star acquisition for this example took around 5 minutes to complete.

We verified star acquisition algorithms acquired a star outside EXCAM FoV and placed the star within the EXCAM capture range. 
During the TVAC tests, other test cases were performed where both dim and bright stars were placed at various locations in the detector's FoV and disturbances were turned ON. For all those cases we verified all EXCAM star acquisition requirements.  All test cases were successful.

\subsection{Raster Scan Star Acquisition tests in TVAC}
 \label{sec:rastertests}

Before raster star acquisition, it was important to calibrate the FSM and LOCAM and complete LOWFSC estimator training activities to find FPM centration axes concerning the CVS axes and set up all preconditions enabling closing the LoS loop, such as closing the FSM inner loop if it was not closed.  It was also important to load an estimator compatible with the star of interest and to configure the LOCAM by computing its gain.

The Raster Scan star acquisition ground command in \textit{iFSW} requires two arguments as input: a raster LOWFSC darksum type and a mask type.   LOWFSC darksum indicates where to access the summed background level on LOCAM.  For TVAC tests, we computed a LOWFSC darksum for each new setup.   Mask types specify whether the raster signal is looking at the reflective light from a disk mask such as NFOV and WFOV or a mask with a bowtie hole that is used for Spectroscopy. An ``HLC NFOV” raster trajectory was set up in all our test cases in TVAC.  We tested raster scan star acquisition using both bright and dim stars and turned on ``flight-like” disturbances during the tests.  We assumed that S/C ACS would place the star within a 0.3 arc-seconds radius of the detector center.    

The objective of these tests was to test \textit{iFSW} and verify that when the star is placed at the required distance from the detector center, FSM rasters the FoV and algorithm can detect the image and find its pixel location and  offload that to the CVS for correction to bring the star to the linear range of LOWFSC so flight software can close the LoS loop autonomously and bring the FSM to its home position. 
 Raster Scan star acquisition must be completed within 5 minutes. Additionally,
the Raster Scan StarID and FindStar algorithms are required to finish within 20 seconds.
  All those requirements were verified, and all test cases were successful.  The next few examples demonstrate CGI raster scan star acquisition in TVAC.

\subsubsection{Raster Scan Star Acquisition Example 1: star at 0.3 arc-seconds radius of detector center }
 \label{sec:rasterex1}
\begin{figure}
    \centering
    \includegraphics[width=0.55\linewidth]{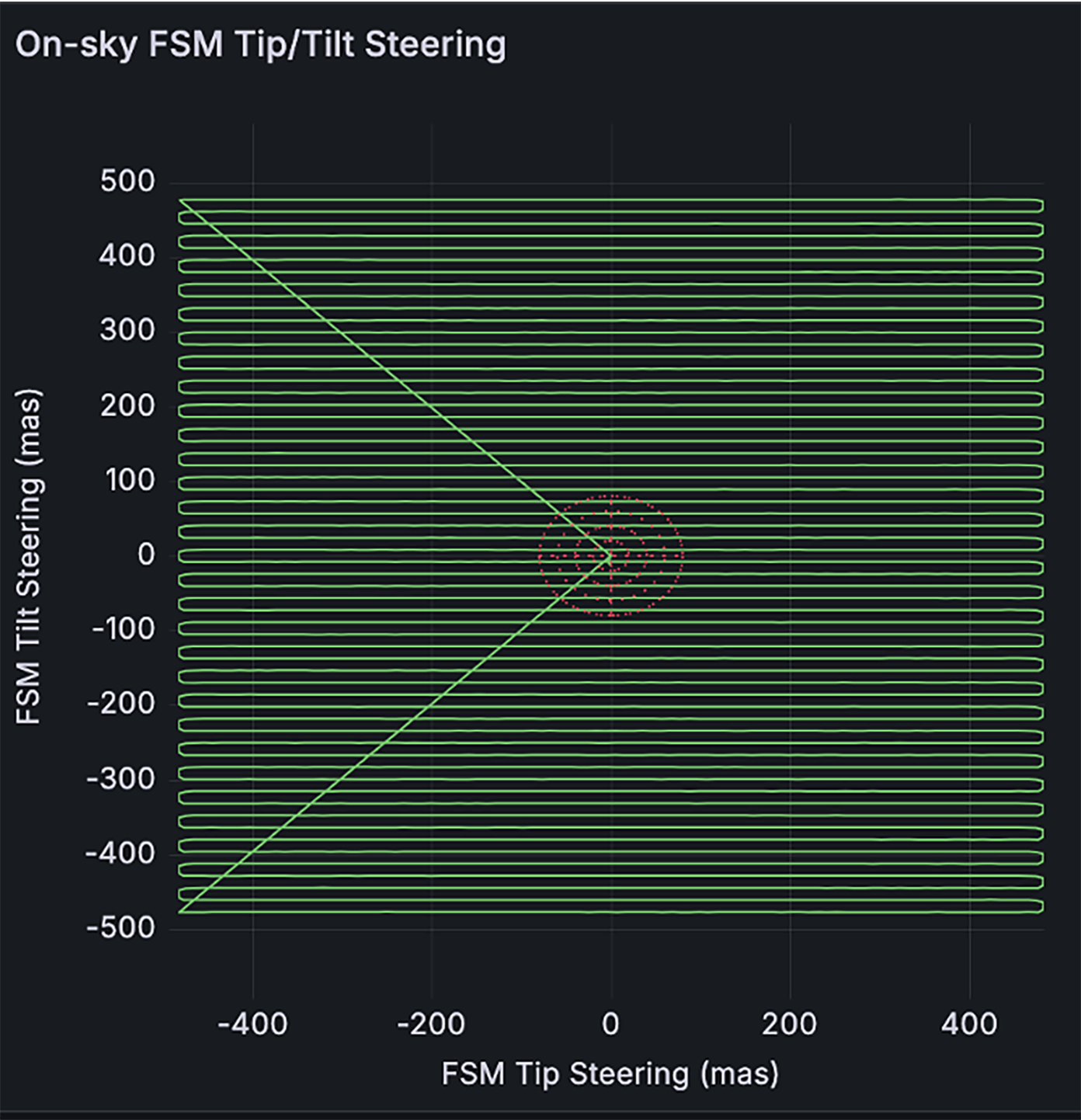}
    \caption{Raster scan star acquisition example 1: The \textit{iFSW} loads the FSM raster trajectory for the HLC mode
    stored in the MRAM. FSM tip/tilt steering where the FSM raster field is 0.48” × 0.48” 
   on the sky and matched the modeled raster trajectory file.}
    \label{fig:Raster_HLC}
\end{figure}

In this example, a star of visual magnitude $2 $ was placed at 0.3 arc-seconds radius from the detector center.  The LOWFSC estimator for this star and LOCAM gain were set up to have all the preconditions to close the LoS loop.  After sending the ground command, the FSM rastered the sky for an HLC mask, as shown in Fig.~\ref{fig:Raster_HLC}.  This verifies that the FSM rasters the sky for an HLC mask as designed. 

\begin{figure}
    \centering
    \includegraphics[width=0.65\linewidth]{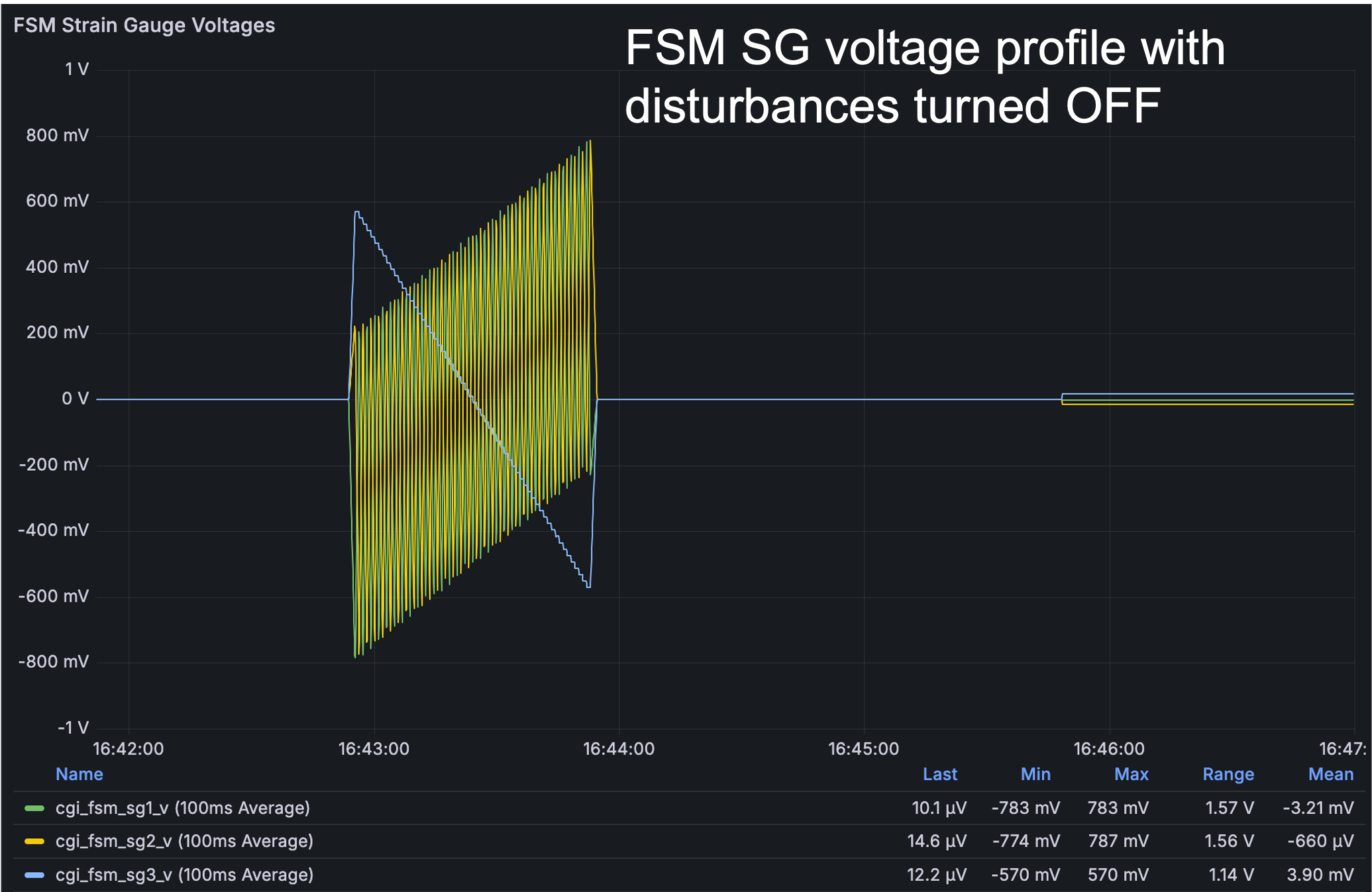}
    \caption{Raster scan star acquisition example 1: The FSM PZTs strain gauge (SG) voltages for all three PZTs
    during the raster scan match the designed strain gauge voltage profiles.}
    \label{fig:FSM_SG_distOff}
\end{figure}

FSM  Strain gauge voltage values during the raster scan created a raster-generated image, and the algorithm determined whether the star was in the FoV covered by the FSM raster. 
Averaged FSM  Strain gauge voltage values for three strain gauges are shown in Fig.~\ref{fig:FSM_SG_distOff} where  FSM strain gauge measurements matched designed values.

\begin{figure}
    \centering
    \includegraphics[width=0.85\linewidth]{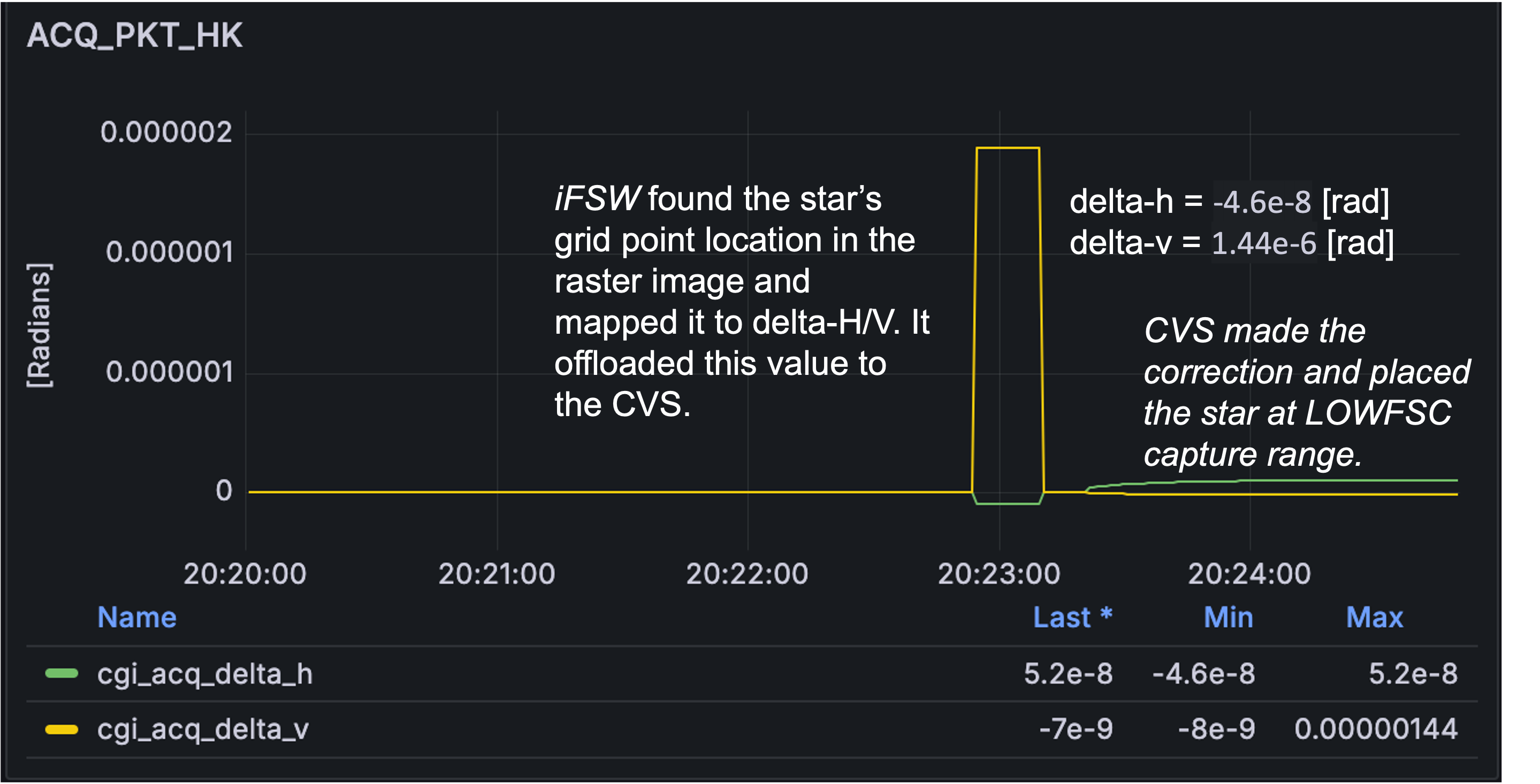}
    \caption{Raster scan star acquisition example 1:  The raster scan acquisition algorithm computed star location
    in the image, and its grid point location was mapped to delta-H/V and was offloaded to CVS for
    correction. The calculated delta-H/V matched the intended star location.}
    \label{fig:Raster_ex1_deltaHV}
\end{figure}

The algorithm then found the star and its grid point location.  It was verified that the grid location matched where the star was placed.  The output PZT voltage values corresponding to the raster grid point were computed and mapped to delta-H and delta-V, as shown in Fig.~\ref{fig:Raster_ex1_deltaHV}.   It was verified that the delta-H/V calculated in the \textit{iFSW} matched the expected values.

\begin{figure}
    \centering
    \includegraphics[width=1\linewidth]{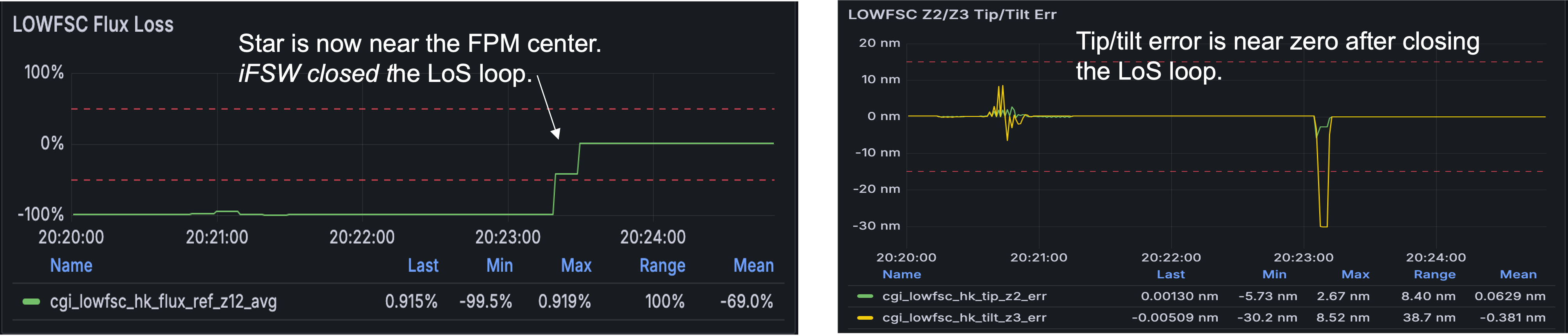}
    \caption{Raster scan star acquisition example 1: The figure shows that the star was placed close to the center
    of the FPM, and the LoS loop was closed at the end of the process.  Z2/Z3 errors are near zero.}
    \label{fig:Raster_ex1_LoS}
\end{figure}

These values were sent to CVS for correction to bring the star within the LOWFSC linear range so the LoS loop could be closed.  Fig.~\ref{fig:Raster_ex1_LoS} shows the star is close to the FPM center where the  \textit{iFSW} closed the LoS loop and Z2/Z3 errors are near zero after closing the LoS loop.   

\begin{figure}
    \centering
    \includegraphics[width=0.35\linewidth]{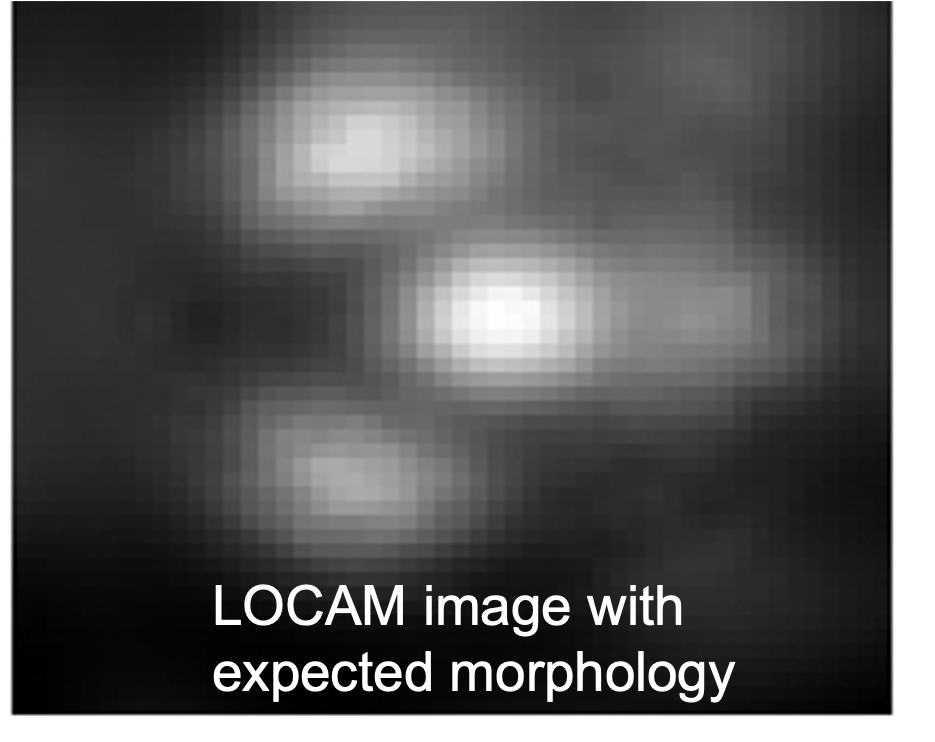}
    \caption{Raster scan star acquisition example 1: LOCAM Image with expected morphology.  This verifies that
    the star is placed at its required position close to the FPM center.}
    \label{fig:Raster_ex1_LOCAM}
\end{figure}

The LOCAM image shown in Fig.~\ref{fig:Raster_ex1_LOCAM} had expected morphology and indicated that the star was close to the FPM center.  

At this point, the CGI tracking state transitioned to “lowfs delta hv”.    
We then received a success message that the raster scan acquisition was successful. 

 As shown in Fig.~\ref{fig:FSM_SG_distOff}, towards the end, after closing the LoS loop, the FSM was centered in its home position, and the FSM strain gauge readings were below the required value.
 
 The whole raster scan star acquisition process took less than the required five minutes duration.  The acquisition algorithms also met the required 20s duration.
 We verified that raster scan star acquisition for a star placed at 0.3 arc-seconds from the detector's center was acquired successfully.

\subsubsection{Raster Scan Star Acquisition Example 2: : star at 0.03 arcseconds radius of detector center }
 \label{sec:rasterex2}
\begin{figure}
    \centering
    \includegraphics[width=0.65\linewidth]{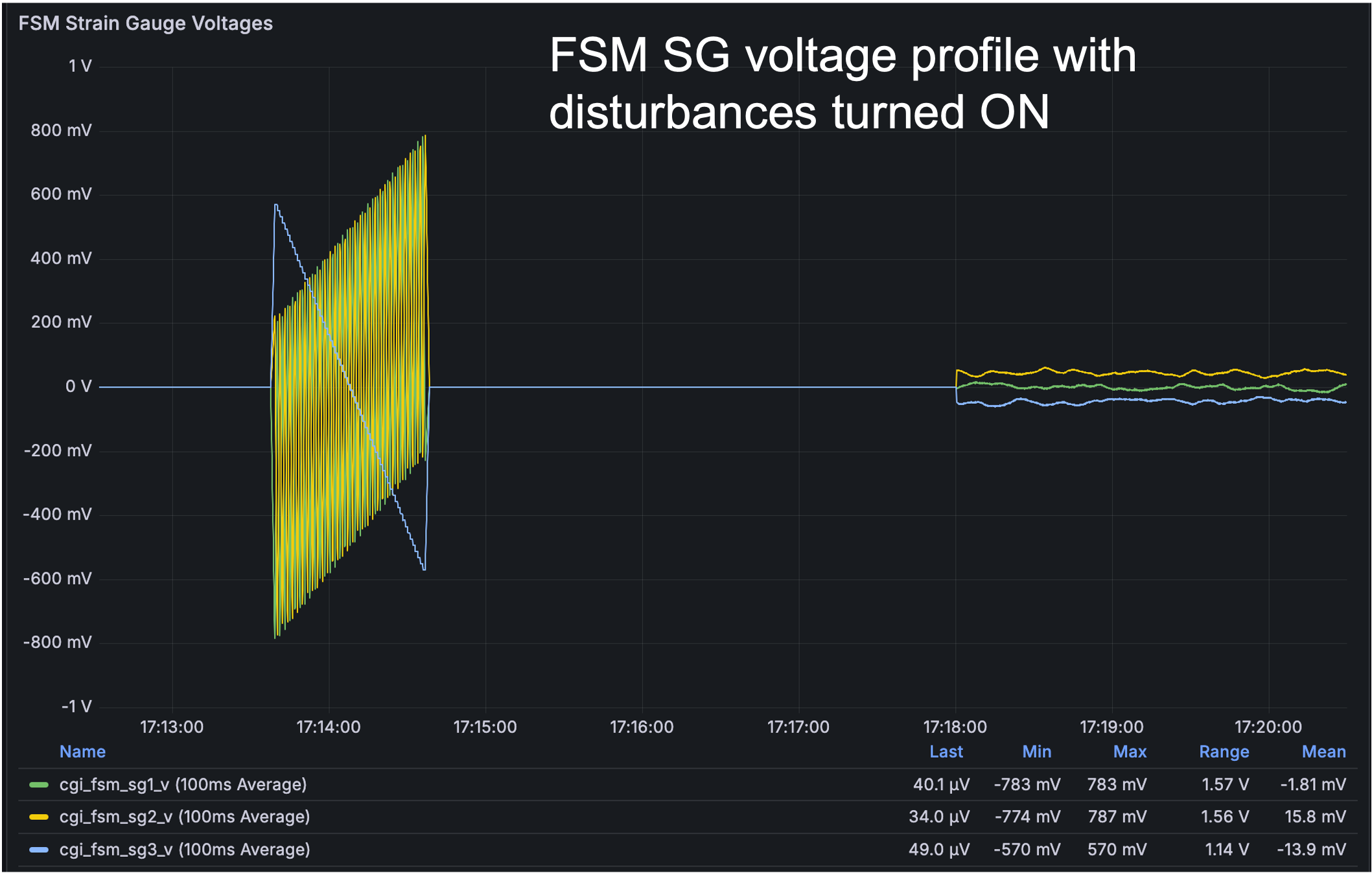}
    \caption{Raster scan star acquisition example 2: The FSM PZTs strain gauge (SG) voltages during the raster
   scan match the designed strain gauge voltage profiles. In this case, the disturbances were turned on.}
    \label{fig:FSM_SG_distOn}
\end{figure}

In this example, a star of visual magnitude 5 was placed at 0.03 arc-seconds radius from the detector center with “flight-like” disturbances turned on.  
The intention for this example was to test raster scan star acquisition of a dim star with disturbances turned ON.  The setup was done like the last example and again the raster scan star acquisition was completed successfully after the LoS loop was closed in the presence of disturbances injected using the CVS jitter mirror.  
FSM  Strain gauge voltage values during the raster scan created a raster-generated image and the algorithm determined whether the star was in the FoV covered by the FSM raster. Averaged FSM  Strain gauge voltage values for 3 strain gauges are shown in Fig.~\ref{fig:FSM_SG_distOn} with disturbances ON and FSM strain gauge measurements matched designed values.

\begin{figure}
    \centering
    \includegraphics[width=0.55\linewidth]{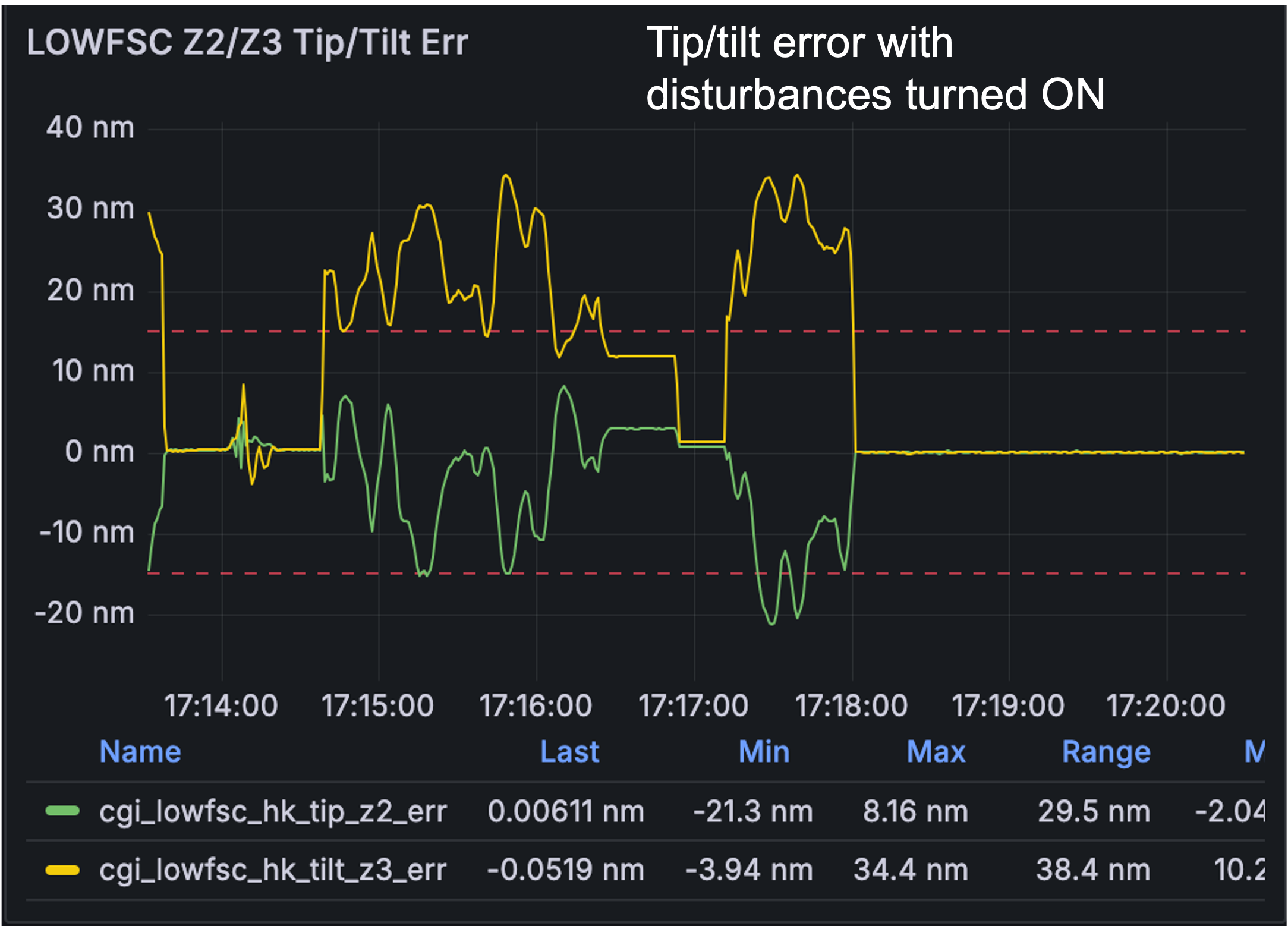}
    \caption{Raster scan star acquisition example 2: Successful raster scan star acquisition in the presence of
     disturbances. \textit{iFSW} closed the LoS loop, resulting in near-zero Z2/Z3 errors.}    \label{fig:RasterEx2_Z2Z3_distOn}
\end{figure}

Fig.~\ref{fig:RasterEx2_Z2Z3_distOn} shows measured Z2/Z3 errors with disturbances on, and after closing the LoS loop at the end of the raster scan star acquisition, these measurements were near zero. This showed that after the star acquisition, the LoS loop was closed and stayed closed. 	
The CGI pointing requirement
(residual  Z2/Z3) after the LoS loop is closed is 1.0 milli-arcsecond RMS per axis in the presence
of sensing noise and external disturbance sources, such as the reaction wheel disturbances,  and
the ACS pointing error.\cite{MM2024}

\begin{figure}
    \centering
    \includegraphics[width=0.30\linewidth]{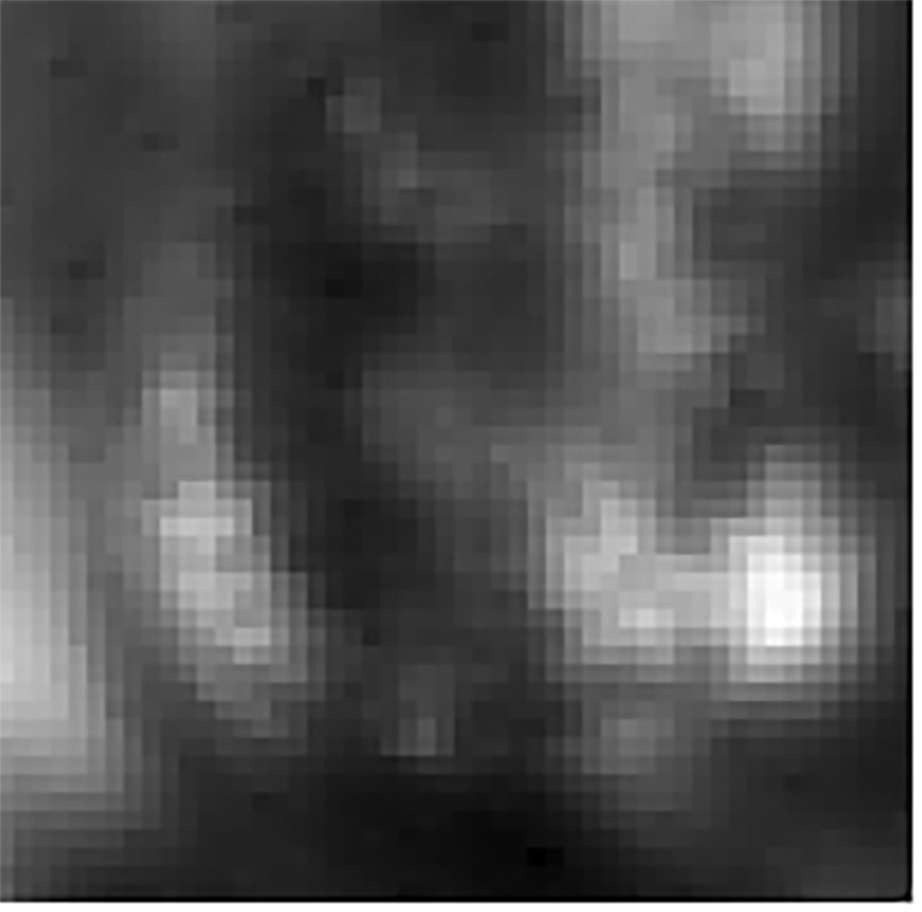}
    \caption{LOCAM image morphology for a failed raster scan star acquisition with LoS loop open.}
    \label{fig:raster_LOS_open}
\end{figure}

Fig.~\ref{fig:raster_LOS_open} shows an example of a failed case when the LoS loop stayed open, and star
acquisition was not completed successfully.  The LOCAM image in Fig.~\ref{fig:raster_LOS_open} did not have the expected morphology as it is expected for a successful raster scan star acquisition shown in Fig.~\ref{fig:Raster_ex1_LOCAM}.

\section{Conclusion and Summary}
Successful star acquisition is one of the key initial tasks CGI must perform during its operational
phase.
CGI star acquisition would employ two different methods depending on mission phases. Both methods rely on the RST Attitude Control System to provide additional pointing capabilities.  This paper presented an overview of the architecture and design of these two methods.

Testing these algorithms at the CGI system level, with their implementation in flight software, has always been a challenge.  Initially, most design verification was conducted through models at earlier stages, and subsequently, after the algorithms were implemented in the flight software, they underwent testing across multiple phases.

A significant part of this challenge lies in the interface between the spacecraft ACS and the CGI during acquisition, requiring close collaboration and well-defined interface requirements between the two systems. Early in the process, the CGI communicated its pointing requirements to the RST ACS, establishing criteria aimed at ensuring the success of star acquisition.

During the CGI TVAC testing campaign, we successfully demonstrated that both star acquisition methods, EXCAM and Raster Scan were performed successfully, and all requirements were met.  We conducted many test cases where stars were placed at the edge or even beyond the detector's field of view, with bright and dim stars for each method and all test cases were successful.  

The CGI testing of these algorithms in earlier stages and across multiple phases proved crucial for this success.
With limited time during CGI TVAC testing, CGI star acquisition was only tested for HLC mode.  Nevertheless, these tests have provided us with confidence that star acquisition is functioning as expected.

During TVAC  tests we operated without the S/C ACS in the loop and used CVS to emulate the ACS functionality.  
The first time we will conduct system-level star acquisition with the ACS in the loop; will be during in-orbit commissioning. However,  both subsystems have maintained communication, exchanging independent models, and validating the approach through simulation and analysis. Additionally,  transformation matrices are available as settable parameters to address any potential phasing and directionality issues.

\subsection* {Disclosures}
The authors declare that there are no financial interests, commercial affiliations, or other potential
conflicts of interest that could have influenced the objectivity of this research or the writing of this
paper.
  
\subsection* {Code, Data, and Materials Availability}
Unless otherwise specified herein for results in star acquisition TVAC tests, the input data and codes used to produce the presented results are not publicly available due to proprietary and export control constraints.

\subsection* {Acknowledgments}
The work described in this paper was carried out at the Jet Propulsion Laboratory, California Institute of Technology, under contract with the National Aeronautics and Space Administration.  The authors would like to thank all CGI Pointing, Acquisition, and Control Element (PACE) team members for contributing to this work.  We also would like to acknowledge TVAC testing campaign team members' contributions, specifically Caleb Baker, Eric Cady, Jason Hauss, Katie Heydorff, and Brian Kern.

\subsection* {Disclosure}
The authors declare that there are no financial interests, commercial affiliations, or other potential conflicts of interest that could have influenced the objectivity of this research or the writing of this paper.

\bibliography{report}   
\bibliographystyle{spiejour}   

\vspace{2ex}\noindent\textbf{Dr. Nanaz Fathpour} is a member of the Guidance \& Control section at Jet Propulsion Laboratory, California Institute of Technology in Pasadena, CA.  She earned her master's degree in applied mathematics and electrical engineering and a Ph.D. in Electrical Engineering (Control theory) from the University of Southern California. During the CGI development, she has been the Pointing, Acquisition, and Control Element (PACE) team lead.  The team was responsible for architecting, designing, analyzing, and delivering code for the CGI star acquisition, pointing LOS control system, Focus, and Zernike control systems. She has contributed to various JPL missions, including the Europa Clipper,  launched in 2024.

\vspace{2ex}

\listoffigures

\end{spacing}
\end{document}